\documentclass{article}

\usepackage{microtype}
\usepackage{graphicx}
\usepackage{subcaption}
\usepackage{booktabs} 
\usepackage{multirow}

\usepackage{hyperref}

\usepackage[preprint]{icml2026}

\usepackage{amsmath}
\usepackage{amssymb}
\usepackage{mathtools}
\usepackage{amsthm}

\usepackage[capitalize,noabbrev]{cleveref}

\theoremstyle{plain}

\theoremstyle{definition}

\theoremstyle{remark}

\usepackage[textsize=tiny]{todonotes}

\icmltitlerunning{Submission and Formatting Instructions for ICML 2026}

\begin{document}

\twocolumn[
  \icmltitle{Co-Generative De Novo Functional Protein Design}



  \icmlsetsymbol{equal}{*}




\begin{icmlauthorlist}
    \icmlauthor{Xinrui Chen}{air,cs}
    \icmlauthor{Yizhen Luo}{air,cs}
    \icmlauthor{Siqi Fan}{air}
    \icmlauthor{Zaiqing Nie$^\dagger$}{air,pharmolix}
\end{icmlauthorlist}

\icmlaffiliation{air}{Institute for AI Industry Research (AIR), Tsinghua University}
\icmlaffiliation{cs}{Department of Computer Science and Technology, Tsinghua University}
\icmlaffiliation{pharmolix}{PharMolix Inc}

\icmlcorrespondingauthor{Xinrui Chen}{cxr21@mails.tsinghua.edu.cn}
\icmlcorrespondingauthor{Zaiqing Nie}{zaiqing@air.tsinghua.edu.cn}

  \icmlkeywords{Machine Learning, ICML}

  \vskip 0.3in
]



\printAffiliationsAndNotice{$^\dagger$Corresponding author.}  


\begin{abstract}


\textit{De novo} functional protein design aims to generate protein sequences that realize specified biochemical functions without relying on evolutionary templates, enabling broad applications in biotechnology and medicine. Existing approaches adopt either direct function-to-sequence mapping or decoupled structure-sequence generation strategies but often fail to achieve functionality and foldability simultaneously.
To address this, we propose \textbf{CodeFP}, a \textbf{Co}-generative protein language model for \textit{\textbf{de} novo} \textbf{F}unctional \textbf{P}rotein design that simultaneously decodes sequence and structure tokens, thereby enabling superior simultaneous realization of functionality and foldability. 
CodeFP utilizes functional local structures to enrich functional semantic encodings, overcoming the suboptimal translation of flat encodings into structure tokens, while introducing auxiliary functional supervision to alleviate training ambiguity stemming from the one-to-many structure-to-token mapping. Extensive experiments show that CodeFP consistently achieves average improvements of 6.1\% in functional consistency and 3.2\% in foldability over the strongest baseline.

\end{abstract}

\section{Introduction}

Functional protein design aims to engineer novel sequences with tailored biological functions, enabling the diverse creation of enzymes with enhanced catalytic efficiency \cite{enzyme_app_1,enzyme_app_2,ZymCTRL}, therapeutic proteins with low toxicity \cite{therapeutic_app1,l21}, and antibodies with improved binding specificity \cite{antibody_app1}. 
Recently, \textit{de novo} functional protein design has attracted increasing interest in biological research \cite{denovo_1,kortemme2024novo}.
Unlike traditional approaches that optimize existing wild-type proteins via directed evolution \cite{dl1,de1,dl2}, \textit{de novo} design operates beyond the space of naturally occurring sequences, thereby jumping out of local fitness optima and enabling novel combinations of multiple functions within a single protein.

\begin{figure}[t] 
    \centering
    \includegraphics[width=\linewidth]{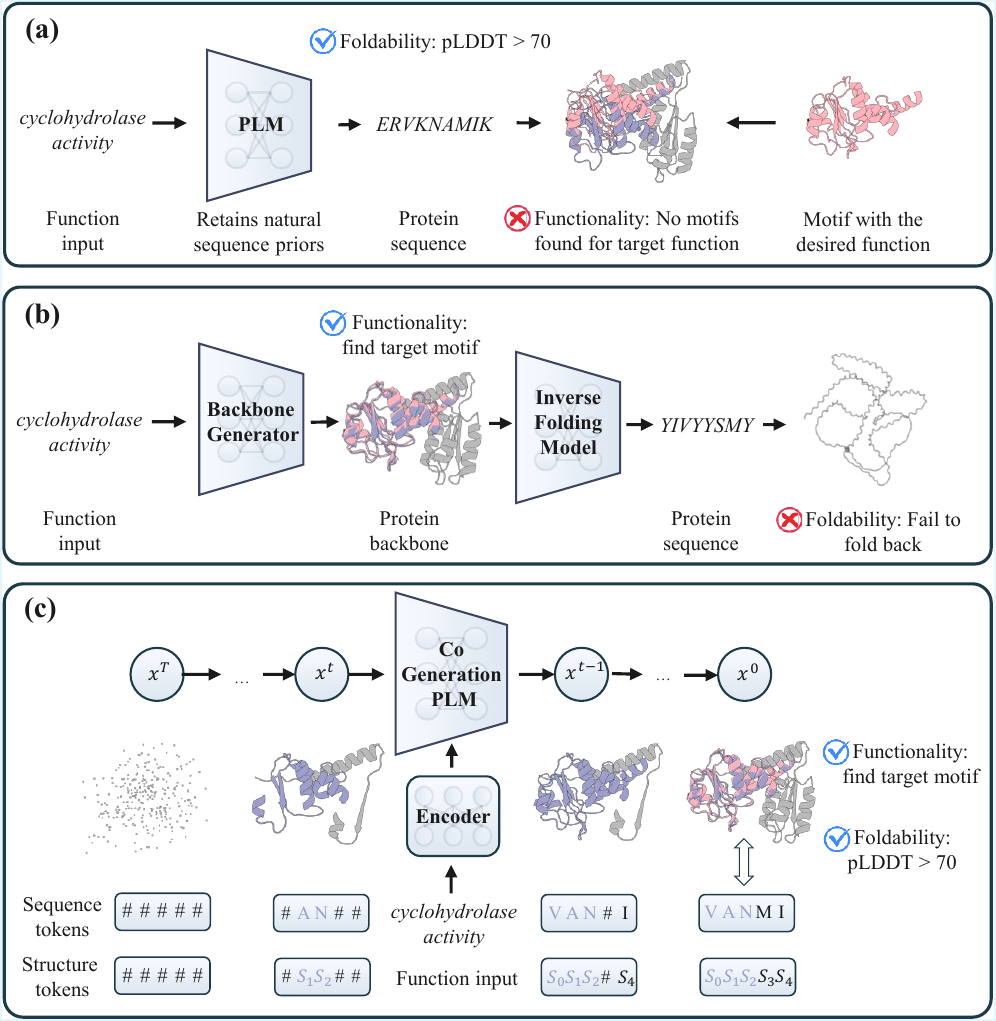}
    
    \caption{\textbf{Motivation of CodeFP.} \textbf{(a)} One-step generation (limited functional control); \textbf{(b)} Two-step generation (unreliable foldability); \textbf{(c)} CodeFP (joint sequence-structure decoding). By iteratively generating both sequence and structure tokens, CodeFP ensures that the generated proteins possess valid folds while retaining critical functionality.}
    \label{fig:why_co_gen}
    \vspace{-5pt}
\end{figure}

Recent advances in machine learning have attempted to address \textit{de novo} functional protein design as a conditional generation task, employing Gene Ontology (GO) terms \cite{go} or natural language to model the desired function. These methods could be generally categorized as follows: 
(1) One-step generation \cite{Progen,ZymCTRL,cfpgen,ProDVa} leverages autoregressive or diffusion-based pre-trained Protein Language Models (PLMs) to map functional conditions directly to amino acid sequences. 
(2) Two-step generation \cite{RFdiffusion,chroma,pinal} incorporates structure as an explicit intermediate modality. Specifically, these models first generate a backbone conditioned on the desired function and then derive the sequence via inverse folding \cite{ProteinMPNN}.

However, due to the intricate coupling among protein sequence, structure, and function, these methods often struggle to generate proteins that simultaneously exhibit \textbf{foldability}\textit{, i.e.,} the sequence should fold into a stable and well-defined three-dimensional structure, and \textbf{functionality}\textit{, i.e.,} the generated protein should exhibit the desired functions.
Specifically, (1) One-step generation, while promoting robust foldability by inheriting natural sequence priors from pre-trained PLMs, often results in degraded functionality due to the diverse sequence realizations underlying a given function that complicates learning, as illustrated in Fig.~\ref{fig:why_co_gen}(a).
(2) Two-step generation, while explicitly modeling structure to ground function, leads to suboptimal foldability as it neglects sequence constraints during backbone generation, yielding geometries that are incompatible with folding back into a natural sequence, as illustrated in Fig.~\ref{fig:why_co_gen}(b).

\begin{figure*}[t] 
    \centering
    \includegraphics[width=\linewidth]{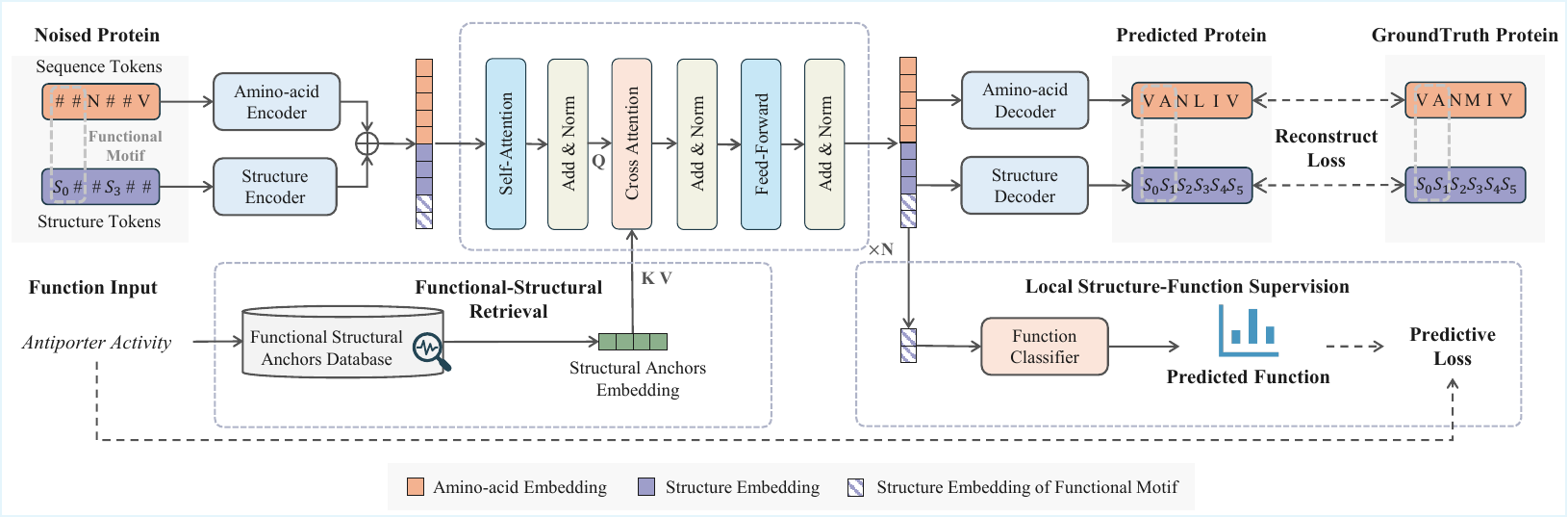}
    
    \caption{\textbf{The overall architecture of CodeFP.} CodeFP facilitates \textit{de novo} functional protein design through a co-generation process. Given a function prompt, the Functional-Structural Retrieval module retrieves representative structural motifs as informative priors. These priors guide the Co-generation Transformer to iteratively reconstruct sequence and structure tokens via cross-attention. In parallel, the Local Structure-Function Supervision module provides auxiliary training signals by classifying the embedding of generated functional local structures. The entire system is trained by minimizing a joint objective of reconstruction and predictive losses.}
\label{fig:architecture}
\end{figure*}

In light of recent advances in co-generative PLMs \cite{dplm2,esm3,coflow}, we introduce \textbf{CodeFP}, a novel \textbf{Co}-generative PLM framework for \textbf{de} novo \textbf{Fun}ctional protein design. CodeFP quantizes local structures for each amino acid into discrete tokens and models them jointly with the protein sequence. During generation, the two modalities are decoded in an interleaved manner, thereby enhancing function modeling via structural integration and ensuring foldability by incorporating sequence constraints, as illustrated in Fig.~\ref{fig:why_co_gen}(c).

Notably, we observe two technical challenges when extending this strategy to \textit{de novo} functional protein design.
First, following prior work \cite{cfpgen,pinal} that encodes functions with one-hot vectors or natural language embeddings and translates them into structure tokens is suboptimal, as it overlooks the hierarchical structure of protein functions and the intricate connections between functions and proteins. Inspired by motif scaffolding \cite{scaffolding1}, we retrieve and encode functional structural motifs. These representations are aggregated by functional category and subsequently integrated via cross-attention to condition the generative process, enhancing the translation from function terms to proteins.
Second, since structural tokenization is sensitive to global topology, functional motifs exhibit diverse realizations in structural token sequences. However, the training objective of discrete diffusion treats them as competing modes, leading to ambiguity. To mitigate this, we apply a functional prediction head to the continuous hidden states of generated local structural motifs as an auxiliary training signal, facilitating function-conditioned learning.










Extensive experiments demonstrate that CodeFP achieves superior functionality and foldability compared to state-of-the-art methods. Quantitatively, it yields a 7.6\% gain in functional F1-Macro and improves the foldability success rate (pLDDT $>70$) by 5.2\% over the strongest baseline. Notably, a 9.1\% improvement in F1-Macro in the out-of-distribution (OOD) test set indicates that CodeFP possesses superior generalization capabilities for unseen functional combinations. Our contributions are summarized as follows:

\begin{itemize}


    \item We propose CodeFP, a co-generative PLM framework for \textit{de novo} functional protein design that effectively satisfies both functionality and foldability.


    \item We aggregate function-specific motifs to capture stronger function semantics, while introducing an auxiliary training signal to mitigate ambiguity arising from structure discretization.



    \item CodeFP achieves the best joint performance in functionality and foldability among all compared methods, establishing a new state-of-the-art in \textit{de novo} functional protein design. 

\end{itemize}

\section{Related Work}


\textbf{Protein Generative Models.} Generative approaches for protein design can be categorized into three paradigms based on their modeling modalities. 
(1) Sequence generation models the probability distribution of amino acids to capture evolutionary patterns. Early approaches, including ProtGPT2 \cite{ProtGPT2} and Prollama \cite{prollama}, employ autoregressive language models to generate protein sequences. In contrast, recent discrete diffusion models like DPLM \cite{dplm} and EvoDiff \cite{evodiff} formulate protein generation as an iterative denoising process.
(2) Structure generation focuses on constructing valid backbone geometries. These methods typically model continuous 3D backbone geometries using diffusion or flow-matching frameworks, including RFdiffusion \cite{RFdiffusion} and FoldFlow \cite{flow4backbone}, whereas approaches such as SLM \cite{SLM} generate autoregressively over discretized structural tokens.
(3) Co-generation accommodates these modalities, enforcing sequence–structure consistency during generation. Representative approaches couple both modalities using multi-modal flow matching, as seen in MultiFlow \cite{multiflow}, or employ dual-channel discrete diffusion, utilized by ESM3 \cite{esm3}, and DPLM-2 \cite{dplm2}.
Building on this emerging paradigm, CodeFP leverages FSR and LSFS to further align these modalities with functional constraints, effectively extending co-generation PLMs to the task of \textit{de novo} functional protein design.

\textbf{Functional Protein Design.}
Functional protein design aims to generate protein sequences with specific biological functions. Early evolution-based methods navigated fitness landscapes to optimize sequences derived from natural variants \cite{dl_r1}. In contrast, recent approaches leverage \textit{de novo} generative models to design novel proteins, generally following either a one-step or two-step paradigm.
(1) One-step approaches focus on direct sequence generation conditioned on function. ProteoGAN \cite{proGAN} utilizes GANs to model label-sequence relationships, while ProGen2 \cite{Progen} and ZymCTRL \cite{ZymCTRL} leverage autoregressive PLMs for functional steering. Recently, CFP-Gen \cite{cfpgen} introduced discrete diffusion to satisfy multiple constraints. 
(2) Two-step approaches prioritize backbone generation: Chroma \cite{chroma} and ProDiT \cite{co_gen_go} generate continuous coordinates via diffusion or flow-matching, whereas Pinal \cite{pinal} predicts discrete structural tokens before amino acid design.
In this work, CodeFP simultaneously generates sequence and structure, integrating the strengths of one-step and two-step approaches.

\section{Method}


In this section, we present the model architecture of CodeFP that facilitates the simultaneous achievement of functionality and foldability. We begin by formalizing the problem and introducing the co-generation framework in Section \ref{sec:preliminaries}. Next, Section \ref{sec:fsr} details the retrieval module, which improves the suboptimal translation of flat semantic encodings. Finally, Section \ref{sec:lsfs} describes the auxiliary supervision, which alleviates the training ambiguity caused by discretization.

\subsection{Generating functional proteins with co-generation}
\label{sec:preliminaries}

\textbf{Problem Formulation.}
We formulate \textit{de novo} functional protein design by representing a protein as $\mathcal{P} = (\mathbf{s})$, where $\mathbf{s} = [s_1, \dots, s_L]$ is an amino acid sequence of length $L$ and each residue $s_i \in \mathcal{V}_{\text{seq}}$ is drawn from the 20 standard amino acids, and specifying its target function using GO molecular function terms $c_{\text{GO}}$, which provide hierarchical labels that support general functional descriptions.
The objective is to model the conditional distribution $p(\mathcal{P} \mid c_{\text{GO}})$.

\textbf{Structure Quantization for Discrete Diffusion.}
Following DPLM-2, we extend the protein defination to $\mathcal{P} = (\mathbf{s}, \mathbf{x})$, where $\mathbf{x} \in \mathbb{R}^{L \times 4 \times 3}$ denotes the backbone atom coordinates (N, C$\alpha$, C, O). Then a LFQ-based \cite{lfq} vector-quantized structure tokenizer maps $\mathbf{x}$ to discrete structure tokens $\mathbf{z} = [z_1, \dots, z_L]$ by capturing local structure contexts of each amino acid. Here, $z_i \in \{0, \dots, |\mathcal{V}_{\text{struct}}|-1\}$ where $\mathcal{V}_{\text{struct}}$ is a fixed-size vocabulary set. This results in a unified discrete representation $\mathcal{P}_{disc} = (\mathbf{s}, \mathbf{z})$.

\textbf{Forward Process: Multimodal Absorbing Diffusion.}
We model the joint distribution of $\mathcal{P}_{\text{disc}}$ using discrete diffusion \cite{discrete_diffusion} with an absorbing corruption process. At each step, tokens are progressively replaced by a modality-specific mask token $[\text{MASK}]$. Let $\mathbf{u}^{(t)} = (\mathbf{s}^{(t)}, \mathbf{z}^{(t)})$ denote the state at diffusion step $t \in \{0, \dots, T\}$, where $\mathbf{u}^{(0)}$ is the clean data and $\mathbf{u}^{(T)}$ approaches a fully masked noise distribution. The forward process is a Markov chain $q(\mathbf{u}^{(t)} \mid \mathbf{u}^{(t-1)})$ with independent transitions across positions and modalities.

For any token $u \in {s, z}$, the transition is defined as
\begin{equation}
q(u^{(t)} \mid u^{(t-1)}) = \mathrm{Cat}\left(u^{(t)}; u^{(t-1)} \mathbf{Q}_t \right),
\end{equation}
where the absorbing transition matrix is
\begin{equation}
\mathbf{Q}_t = \mathrm{diag}(1 - \beta_t) + \beta_t \cdot \mathbf{1}{[\text{MASK}]},
\end{equation}
Here, $\beta_t$ controls the corruption rate, and $\mathbf{1}{[\text{MASK}]}$ assigns all probability mass to the absorbing mask state.

\textbf{Reverse Denoising with Functional Conditioning.}
The generative process reconstructs the clean protein $\mathbf{u}^{(0)}$ from the corrupted state $\mathbf{u}^{(t)}$ by reversing the diffusion trajectory conditioned on $\mathcal{C} = \{\mathbf{C}_{GO}\}$, which encodes functional semantics of GO terms. The reverse transition is approximated by marginalizing over the predicted clean state:
\begin{equation}
\begin{split}
    p_\theta(\mathbf{u}^{(t-1)} | \mathbf{u}^{(t)}, \mathcal{C}) \propto \\
    \sum_{\tilde{\mathbf{u}}^{(0)}} q(\mathbf{u}^{(t-1)} | \mathbf{u}^{(t)}, \tilde{\mathbf{u}}^{(0)}&) p_\theta(\tilde{\mathbf{u}}^{(0)} | \mathbf{u}^{(t)}, \mathcal{C}),
\end{split}
\end{equation}
where $p_\theta(\cdot | \mathbf{u}^{(t)}, \mathcal{C})$ denotes the neural network prediction. By sustaining dense mutual interaction at every denoising step, our iterative decoding strategy ensures that structural generation is tightly constrained by sequence constraints. Simultaneously, this progressive refinement grants the structural topology sufficient flexibility to extensively explore the geometric space for functional alignment.

\textbf{Optimization Objective.}
The generative objective $\mathcal{L}_{\text{gen}}$ minimizes the variational lower bound for the joint distribution, which reduces to a weighted sum of negative log-likelihoods. Independent time steps $t_s$ and $t_z$ are sampled for sequence and structure modalities, respectively. The loss is formulated as:
\begin{equation}
\begin{split}
\mathcal{L}_{\text{gen}} &= \mathbb{E}_{q(\mathbf{u}^{(0)})} \bigg[ \sum_{i=1}^{L} \bigg( \lambda(t_s) b_i(t_s) \mathcal{L}_{\text{seq}}^{(i)} \\
&\quad + \lambda(t_z) b_i(t_z) \mathcal{L}_{\text{struct}}^{(i)} \bigg) \bigg],
\end{split}
\end{equation}
where $b_i(t) \in \{0, 1\}$ indicates whether the token at position $i$ is masked at time $t$, and $\mathcal{L}^{(i)}$ represents the negative log-likelihood of the reconstruction.

\begin{table*}[t]
\centering
\caption{\textbf{Main results on GO-conditioned protein design.} We evaluate functionality using the DeepGO-SE classifier. $\uparrow$ indicates higher is better, $\downarrow$ indicates lower is better. The best results among generative models are highlighted in \textbf{bold}, and the second best are \underline{underlined}. Positive Control represents real proteins from the test set.}
\label{tab:main_results}
\resizebox{\textwidth}{!}{%
\begin{tabular}{llccccccc}
\toprule
\textbf{Category} & \textbf{Model} & \textbf{F1-Micro} ($\uparrow$) & \textbf{F1-Macro} ($\uparrow$) & \textbf{AUPR} ($\uparrow$) & \textbf{AUC-ROC} ($\uparrow$) & \textbf{MRR} ($\uparrow$) & \textbf{MMD} ($\downarrow$) & \textbf{MMD-G} ($\downarrow$) \\
\midrule
Reference & Positive Control & 0.543 & 0.522 & 0.402 & 0.775 & 0.939 & 0.000 & 0.000 \\
\midrule
\multirow{3}{*}{One-step} 
 & ProteoGAN & 0.376 & 0.093 & 0.121 & 0.510 & 0.277 & \textbf{0.095} & \textbf{0.055} \\
 & ProGen2   & 0.414 & 0.355 & 0.240 & 0.663 & 0.545 & 0.109 & 0.064 \\
 & CFP-Gen   & 0.429 & \underline{0.370} & \underline{0.245} & \underline{0.674} & \underline{0.601} & 0.112 & \underline{0.060} \\
\midrule
\multirow{2}{*}{Two-step} 
 & Chroma & 0.262 & 0.067 & 0.076 & 0.501 & 0.018 & 0.313 & 0.183 \\
 & Pinal  & \underline{0.452} & 0.369 & 0.229 & 0.663 & 0.379 & 0.223 & 0.131 \\
\midrule
Ours & CodeFP & \textbf{0.496} & \textbf{0.446} & \textbf{0.321} & \textbf{0.724} & \textbf{0.658} & \underline{0.106} & 0.063 \\
\bottomrule
\end{tabular}%
}
\end{table*}

\subsection{Functional-Structural Retrieval}
\label{sec:fsr}



Protein functional semantics exhibit deep dependencies on both sequence and structure. Existing methods \cite{pinal,cfpgen}, which rely on one-hot encodings or textual embeddings, suffer from two critical limitations. First, they neglect the hierarchical context of biological functions, such as ATPase activity, which often necessitate capabilities like electron transport. Second, they suffer from geometric decoupling, ignoring the physical reality that functions like ligand binding or enzymatic catalysis are instantiated by specific structural motifs. 
To address these limitations, we ground functional labels in their physical manifestations. As illustrated in Fig.~\ref{fig:architecture}, our method proceeds in two phases: constructing a retrieval database of functional structural motifs, and injecting these priors via cross-attention.

\textbf{Construction of Functional Structural Representation.}
We construct a retrieval database $\mathcal{M}$ that maps each GO term to a continuous structural embedding representing its geometric realization. This process comprises two steps: Representation Encoding and motif Aggregation.



\textit{Representation Encoding.}
Since specific biological functions are governed by local structural motifs rather than the global fold, accurately modeling function requires isolating its geometric instantiation. To achieve this, we utilize the pre-computed domain terms provided in our training set, derived using InterProScan (IPS) \cite{interproscan}. Based on these terms, we extract the local backbone coordinates $\mathbf{x}_{local}$ corresponding to each protein-GO pair.  To translate this geometry into a functional semantic space, we encode $\mathbf{x}_{local}$ using the frozen DPLM-2 encoder—ensuring alignment with our CodeFP backbone. Specifically, the coordinates are discretized via LFQ and processed to extract the \texttt{[CLS]} representation $\mathbf{e}_{i,j}$. This resulting embedding effectively captures the intrinsic dependency between the function and its underlying local structure.

\textit{Motif Aggregation.} 
A GO term $y$ may be associated with diverse proteins, each carrying evolutionary specificities unrelated to the core function. To distill the essential geometric signature of the function and inject an inductive bias for hierarchical protein function, we compute the structural motif $\mathbf{c}_y$ by averaging all local structure embeddings $\mathbf{e}_{i,y}$ associated with label $y$ (i.e., $\mathbf{c}_y = \text{Mean}(\{\mathbf{e}_{i,y} \mid (P_i, y) \in \mathcal{S}_y\})$). Crucially, aggregation preserves the hierarchical structure of function, since the aggregate representation of a parent function naturally encompasses its child nodes.
This centroids serve as hierarchically aware structural motifs, forming our retrieval database $\mathcal{M} = \{ (y, \mathbf{c}_y) \}_{y \in \mathcal{Y}}$.

\textbf{Injection via Cross-Attention.}
During both training and inference, we inject these structural motifs into the co-generation process. Given a set of input GO labels $\mathcal{Y}_{in}$, we retrieve their corresponding structural motifs $\mathbf{C} = \{ \mathbf{c}_y \mid y \in \mathcal{Y}_{in} \}$. 
These motifs are then fused into the model representation via cross-attention layers, augmenting the conditioning set of the reverse denoising process from $\mathcal{C} = \{\mathbf{C}_{GO}\}$ to $\mathcal{C} = \{\mathbf{C}_{GO}, \mathbf{C}\}$.
Let $\mathbf{H}^{(l)}$ denote the hidden states of the sequence and structure tokens at layer $l$. The injection is formulated as:
\begin{equation}
    \mathbf{H}^{(l)'} = \mathbf{H}^{(l)} + \text{CrossAttn}(\mathbf{Q}=\mathbf{H}^{(l)}, \mathbf{K}=\mathbf{C}, \mathbf{V}=\mathbf{C}),
\end{equation}
where the generated tokens (Query) attend to the retrieved structural motifs (Key/Value). By grounding hierarchical functional knowledge in a structural perspective, we introduce a effective inductive bias, facilitating the learning of functional semantics.

\subsection{Local Structure-Function Supervision}
\label{sec:lsfs}

While co-generative discrete diffusion models effectively capture the joint distribution of sequence and structure, optimizing them for functional constraints remains challenging due to the training ambiguity induced by the quantization discrepancy inherent in structural tokenizers.
Unlike standard approaches that supervise solely on discrete outputs, we apply supervision directly to the CodeFP's continuous hidden states.

\textbf{Formulation.}
During training, let $\mathbf{H}^{(L)} \in \mathbb{R}^{T \times d}$ be the continuous hidden states from the last transformer layer. For a protein annotated with GO label $y$, we first identify the indices of the structure tokens corresponding to the functional domain, using the same IPS-based localization described in Section \ref{sec:fsr}.
To facilitate downstream classification, we aggregate the hidden states at the relevant indices via mean pooling to obtain a continuous proxy for the functional domain's structure. We then employ a parameterized classifier head to project this embedding directly into the functional label space, yielding the logits corresponding to the $C$ GO terms. 
The classifier head and CodeFP are optimized jointly, a process facilitated by the frozen model decoder which ensures that the learned embeddings remain aligned with the distribution of natural proteins.

\textbf{Class-Imbalanced Optimization.}
To mitigate the impact of the long-tailed distribution in functional terms, we employ a mean-normalized inverse class frequency strategy. Specifically, we assign a scaling factor $w_c = N_c^{-1} / (\frac{1}{C} \sum_{j} N_j^{-1})$, where $N_i$ means the training set frequency for GO $i$, to balance the dominance of head classes. The final objective is minimized via weighted cross-entropy, defined as $\mathcal{L}_{\text{LSFS}} = - w_y \log \hat{p}_y$ for a target class $y$.

\textbf{Total Training Objective.}
The derived auxiliary loss $\mathcal{L}_{\text{LSFS}}$ is integrated into the generative objective $\mathcal{L}_{\text{gen}}$ (defined in Equation 4). The total optimization objective is thus formulated as:
\begin{equation}
\mathcal{L}_{\text{total}} = \mathcal{L}_{\text{gen}} + \gamma \mathcal{L}_{\text{LSFS}},
\end{equation}
where $\gamma$ serves as a balancing coefficient that scales the gradient contribution of the functional supervision. By imposing supervision on latent space, CodeFP facilitates the incorporation of precise functional supervision signals, alleviating the training ambiguity.

\section{Experiment}

\subsection{Experiment Setup}

\textbf{Dataset.} We adopt the dataset collected by \citet{cfpgen}, which comprises 103.9K protein sequences annotated with 375 GO terms derived from SwissProt \cite{uniprot2025uniprot} and InterPro \cite{blum2025interpro}. The dataset is split into 95.6K training, 831 validation, and 8.3K testing samples. The test set contains 435 unique GO label combinations, including 76 combinations not observed during training. To facilitate structure-sequence co-generation, we first retrieve the pre-computed structure tokens of $\sim$ 50K sequences that overlap with DPLM-2's \cite{dplm2} training set. Then, we query the PDB \cite{pdb} and AlphaFoldDB \cite{afdb} databases to obtain the experimentally resolved or predicted 3D structures of remaining proteins and perform tokenization using DPLM-2's pre-trained LFQ encoder. We filter out 1.4K samples without a publicly available structure.  

\textbf{Implementation Details.} We initialize CodeFP from the pre-trained DPLM-2 (650M) and apply cross-attention to each layer of the Transformer block. We train the cross-attention modules and the LSFS prediction head while keeping the remaining parameters frozen. During sampling, we draw the length of the protein uniformly between 200 and 400, and adopt the same procedure as DPLM-2 to obtain the structure tokens and amino acid tokens using 500 diffusion steps. The model is trained for approximately 60 epochs, taking about 48 hours and achieves an inference latency of about one minute per protein on a NVIDIA A800 GPU. Detailed hyperparameter settings are provided in Appendix~\ref{app:hypermeter}.

\begin{table}[t]
\centering
\caption{\textbf{Foldability evaluation.} We report the structural success rates predicted by ESMFold.}
\label{tab:foldability}
\resizebox{0.9\linewidth}{!}{%
\begin{tabular}{lcc}
\toprule
\textbf{Model} & \textbf{pLDDT $>$ 70 (\%)} & \textbf{pTM $>$ 0.5 (\%)} \\
\midrule

Chroma & 23.47 & 66.76 \\
CFP-Gen & \underline{75.52} & 72.30 \\
Pinal & 74.22 & \underline{82.22} \\
\midrule
CodeFP (Ours) & \textbf{80.65} & \textbf{83.48} \\
\bottomrule
\end{tabular}%
}
\end{table}


\textbf{Baselines.} We benchmark against five representative methods spanning two paradigms:   
(1) One-step generation, including ProteoGAN \cite{proGAN} that adopts a conditional GAN as the backbone, ProGen2 \cite{Progen} that leverages an autoregressive PLM to generate the sequence, and CFP-Gen \cite{cfpgen} that performs discrete diffusion on amino acid sequences. 
(2) Two-step generation, including Chroma \cite{chroma} that uses continuous diffusion and Pinal\cite{pinal} that generate discrete structure tokens.

\textbf{Metrics.} We evaluate our model across three primary dimensions: functionality, foldability, and generative distribution, following \cite{cfpgen}.
(1) \textit{Functionality}: We assess functional fidelity from two perspectives. First, we employ Mean Reciprocal Rank (MRR) that directly evaluates the sequence similarity between generated proteins and ground-truth functional analogs. Second, we apply DeepGO-SE \cite{deepgo} to predict GO terms based on generated sequences. We compare the predicted GO terms with the desired functions and report F1-Micro, F1-Macro, AUPR, and AUC-ROC scores. We also report Exact and Partial Match rates, which quantify whether all or any of the desired functions are recovered based on DeepGO's predictions. 
(2) \textit{Foldability}: To assess whether sequences adopt stable, physically realizable conformations, we employ ESMFold~\cite{esm2_esmfold} for structure prediction. A sequence is considered structurally successful if it achieves a mean pLDDT score above 70, indicating reliable local confidence, and a pTM score above 0.5, reflecting a consistent global fold.
(3) \textit{Generative Distribution}: We evaluate the generative distribution with diversity, novelty, and adherence to the natural protein distribution. Diversity captures variability among generated sequences and is defined as one minus the mean pairwise sequence identity. Novelty quantifies dissimilarity from the training set and is computed as one minus the maximum sequence identity to any training protein using MMseqs2~\cite{mmseqs2}; higher values indicate better performance for both metrics. In addition, we assess distributional alignment with natural protein sequences using Maximum Mean Discrepancy (MMD) and its Gaussian-kernel variant (MMD-G), where lower values denote closer alignment. Implementation details are provided in Appendix~\ref{app:metrics}.    

\subsection{Main Results}

\paragraph{CodeFP achieves superior functionality performance.}
As shown in Table~\ref{tab:main_results}, our model surpasses Pinal in F1-micro (0.496 vs. 0.452) and outperforms CFP-Gen in both AUC-ROC (0.724 vs. 0.674) and MRR (0.658 vs. 0.601). These improvements indicate that the generated proteins more accurately capture functional specificity while reducing spurious assignments, and better align with functional analogs observed in natural proteins, highlighting the model’s ability to capture intrinsic relationships between functional semantics and protein sequences.

\paragraph{Improved coverage of long-tailed functional categories.} 
The performance gap widens on imbalance-sensitive metrics, with substantial gains over CFP-Gen in both F1-macro (0.446 vs. 0.370) and AUPR (0.321 vs. 0.245), indicating improved modeling of long-tailed functional categories. In contrast to baselines that tend to favor high-frequency functional modes, our method exhibits effective generalization to the long-tailed distribution, consistent with the complementary effects of FSR in facilitating functional abstraction and LSFS in reducing training ambiguity.

\paragraph{State-of-the-art foldability of generated proteins.} 
As reported in Table~\ref{tab:foldability}, our model achieves the highest success rates surpasses Pinal in both pLDDT (80.65\% vs. 74.22\%) and pTM (83.48\% vs.\ 82.22\%). 
These results indicate that the generated proteins are more likely to fold into well-defined structures, consistent with the benefits of jointly modeling sequence and structure.

\begin{figure}[t] 
    \centering
    \includegraphics[width=\linewidth]{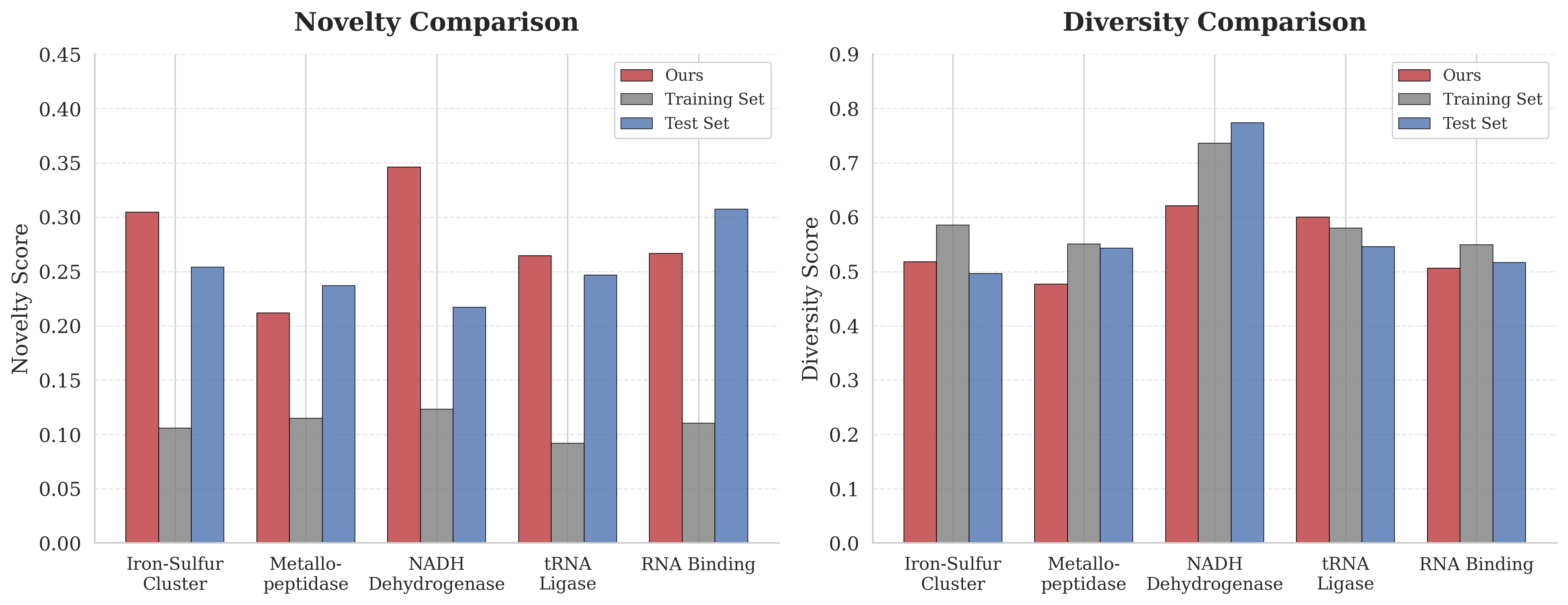}
    
    \caption{\textbf{Analysis of generative novelty and diversity.} We illustrate the distribution of Novelty (left) and Diversity (right) across five diverse functional tasks.}
    \label{fig:novelty_diversity}
\end{figure}

\paragraph{Preservation of natural protein distribution.}
Consistently, all one-step generation models exhibit comparable MMD scores, whereas two-step approaches suffer substantially worse distributional alignment.
Our model achieves competitive MMD (0.106 vs.\ 0.095) and MMD-G (0.063 vs.\ 0.055) against ProteoGAN, indicating that improved functional controllability does not compromise alignment with the natural protein distribution. We attribute this performance to the sequence priors inherited from large-scale sequence pretraining. 

\begin{figure}[t]
    \centering
    \includegraphics[width=\linewidth]{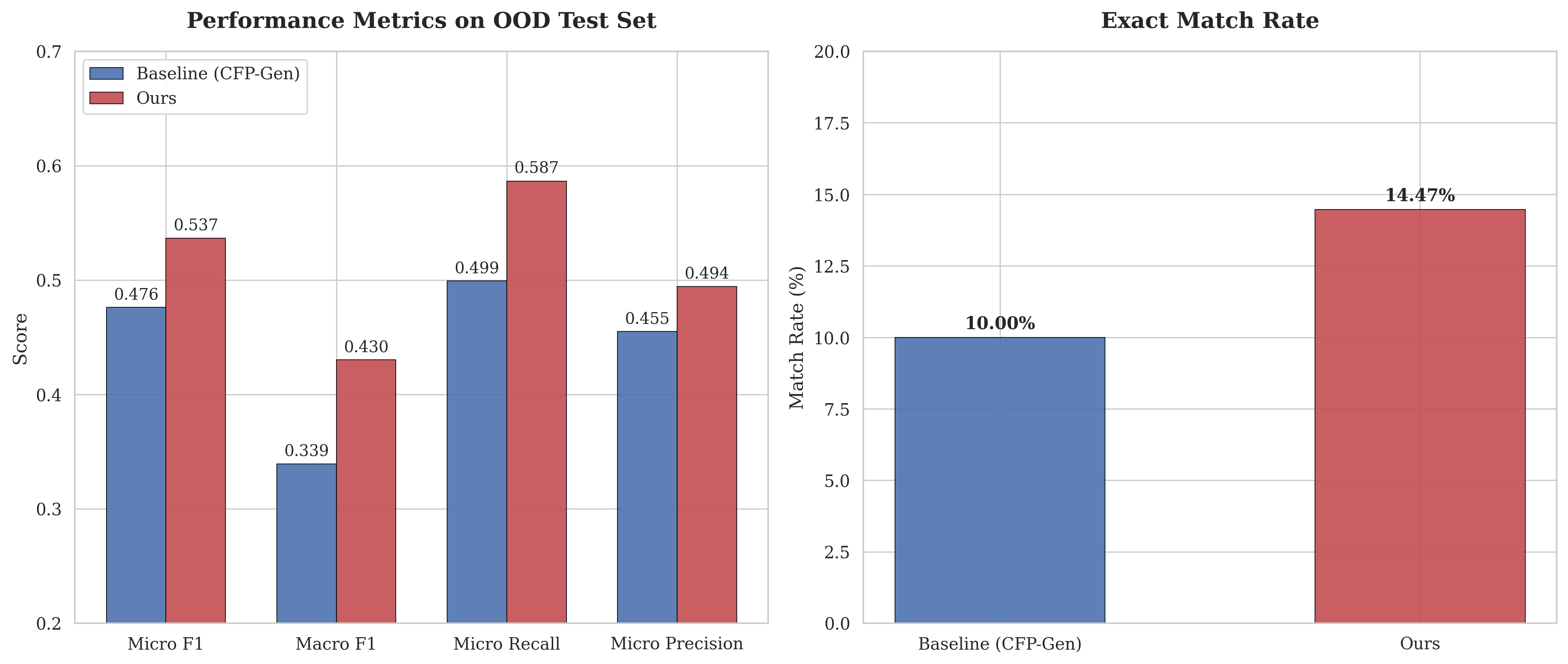}
    \caption{\textbf{Performance on OOD functional combinations.} We report multi-label classification metrics and the exact match rate on OOD test subset.}
    \label{fig:ood_metrics}
    \vspace{-0pt}
\end{figure}

\paragraph{Novelty and diversity.}
To assess the ability of our model to generate novel yet functional protein sequences under a constrained inference budget, we analyze five functional combinations spanning diverse biological mechanisms. For each combination, we generate 30 sequences and compare them with equal-sized samples drawn from the training and test sets. 
As shown in Fig.~\ref{fig:novelty_diversity} (left), our model consistently achieves novelty scores substantially higher than the training set and generally comparable to those of the natural test set, indicating that CodeFP leverages retrieved results as functional priors to explore new regions of the sequence space.
Meanwhile, Fig.~\ref{fig:novelty_diversity} (right) shows that the diversity of the generated sequences closely matches the natural diversity observed in both training and test sets, indicating sufficient exploration of the sequence space.

\begin{table*}[t]
\centering
\caption{\textbf{Ablation study on component contributions.} We analyze the impact of co-generation, FSR, and LSFS on functional consistency, distributional alignment, and foldability.}
\label{tab:ablation}
\resizebox{0.95\textwidth}{!}{%
\begin{tabular}{lcccccc}
\toprule
\multirow{2}{*}{\textbf{Model Variant}} & \multicolumn{4}{c}{\textbf{Functional Consistency \& Distribution}} & \multicolumn{2}{c}{\textbf{Structural Realizability}} \\
\cmidrule(lr){2-5} \cmidrule(lr){6-7}
 & \textbf{F1-Micro} ($\uparrow$) & \textbf{F1-Macro} ($\uparrow$) & \textbf{MRR} ($\uparrow$) & \textbf{MMD} ($\downarrow$) & \textbf{pLDDT $>$ 70} (\%) & \textbf{pTM $>$ 0.5} (\%) \\
\midrule
CodeFP & \textbf{0.496} & \textbf{0.446} & \underline{0.658} & \underline{0.106} & \underline{80.65} & \textbf{83.48} \\
\midrule
w/o LSFS & \underline{0.495} & \underline{0.437} & 0.645 & 0.172 & \textbf{82.01} & \underline{81.73} \\
w/o FSR & 0.486 & 0.423 & \textbf{0.674} & \textbf{0.101} & 71.57 & 76.76 \\
w/o FSR \& LSFS & 0.465 & 0.400 & 0.534 & 0.192 & 71.71 & 71.98 \\
two-step generation & 0.414 & 0.285 & 0.312 & 0.282 & 52.24 & 59.54 \\

\bottomrule
\end{tabular}%
}
\end{table*}

\subsection{Generalization to Out-of-Distribution Functional Combinations}

We evaluate model generalization across two Out-of-Distribution (OOD) scenarios: (1) Unseen Natural Combinations, which occur in nature but are withheld from the training data, and (2) Hypothetical Combinations, which violate natural co-occurrence patterns and have no known biological instances.

\textbf{Unseen Natural Combinations.}
We curate a test set of 76 functional combinations held out during training, generating 10 candidates per combination for assessment. 
As shown in Fig.~\ref{fig:ood_metrics}, the low Exact Match Rates (peaking at only 14.5\%) highlight the inherent difficulty of precisely realizing novel functional pairings. Nevertheless, the relatively high F1 and Recall scores indicate that the model captures partial functional constraints. Despite these challenges, our model outperforms the baseline by 9.1\% in F1-Macro and 4.47\% in Exact Match Rate, demonstrating superior zero-shot synthesis capabilities.

\begin{figure}[t]
    \centering
    \includegraphics[width=\linewidth]{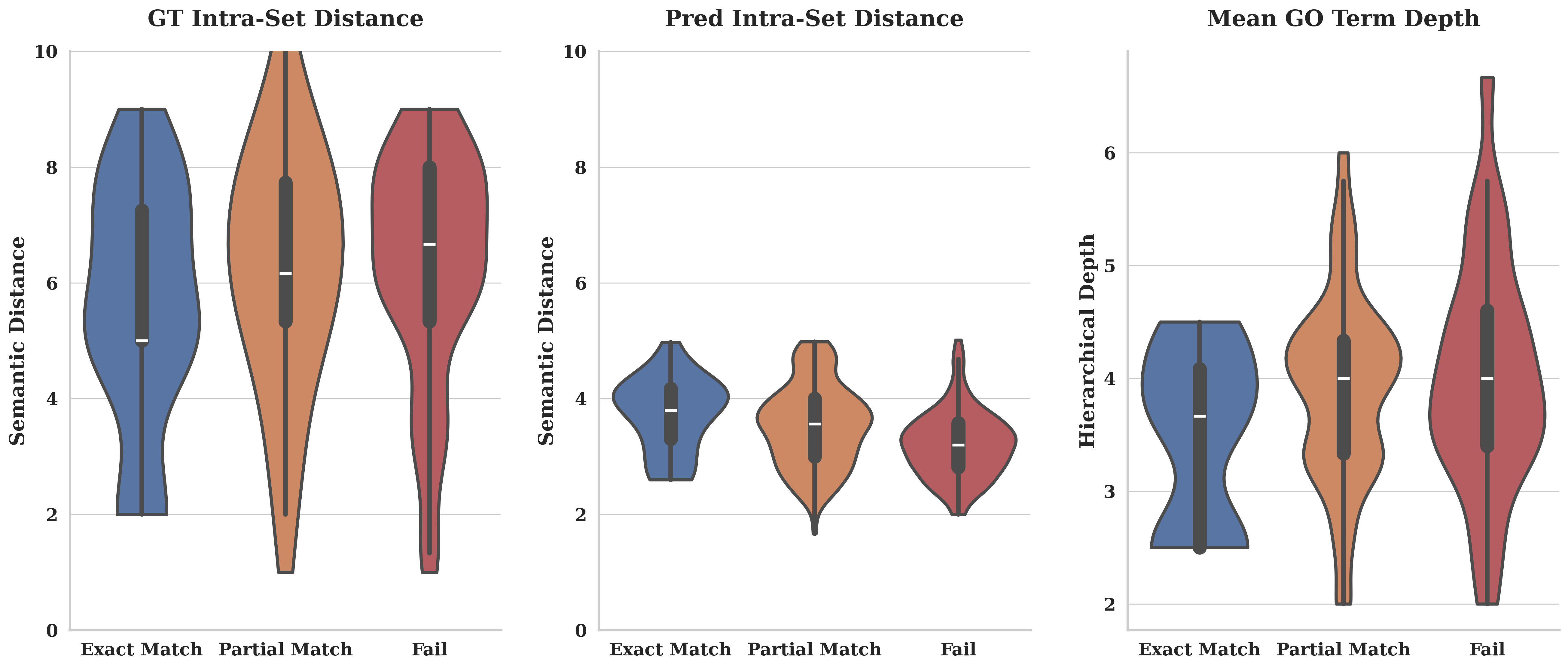} 
    \caption{\textbf{Attribute analysis of generation difficulty.} We analyze generation outcomes against three topological attributes of the GO graph.}
    \label{fig:ood_attributes}

    \vspace{-15pt}
\end{figure}

To further elucidate the mechanisms underlying OOD generalization, we analyze model performance with respect to GO graph topology (Fig.~\ref{fig:ood_attributes}). Specifically, semantic distance measures functional dissimilarity between GO terms, while term depth captures functional specificity, with deeper terms corresponding to more specialized functions.
We observe that successful generations are associated with smaller semantic distances between target terms, indicating that functional similarity facilitates compositional synthesis. In contrast, failures are linked to greater GO term depth, suggesting that highly specific functions are more difficult to integrate. Moreover, these failures exhibit reduced predicted semantic diversity, reflecting a collapse toward a narrower and functionally homogeneous semantic space.

\begin{figure}[t]
    \centering
    \includegraphics[width=\linewidth]{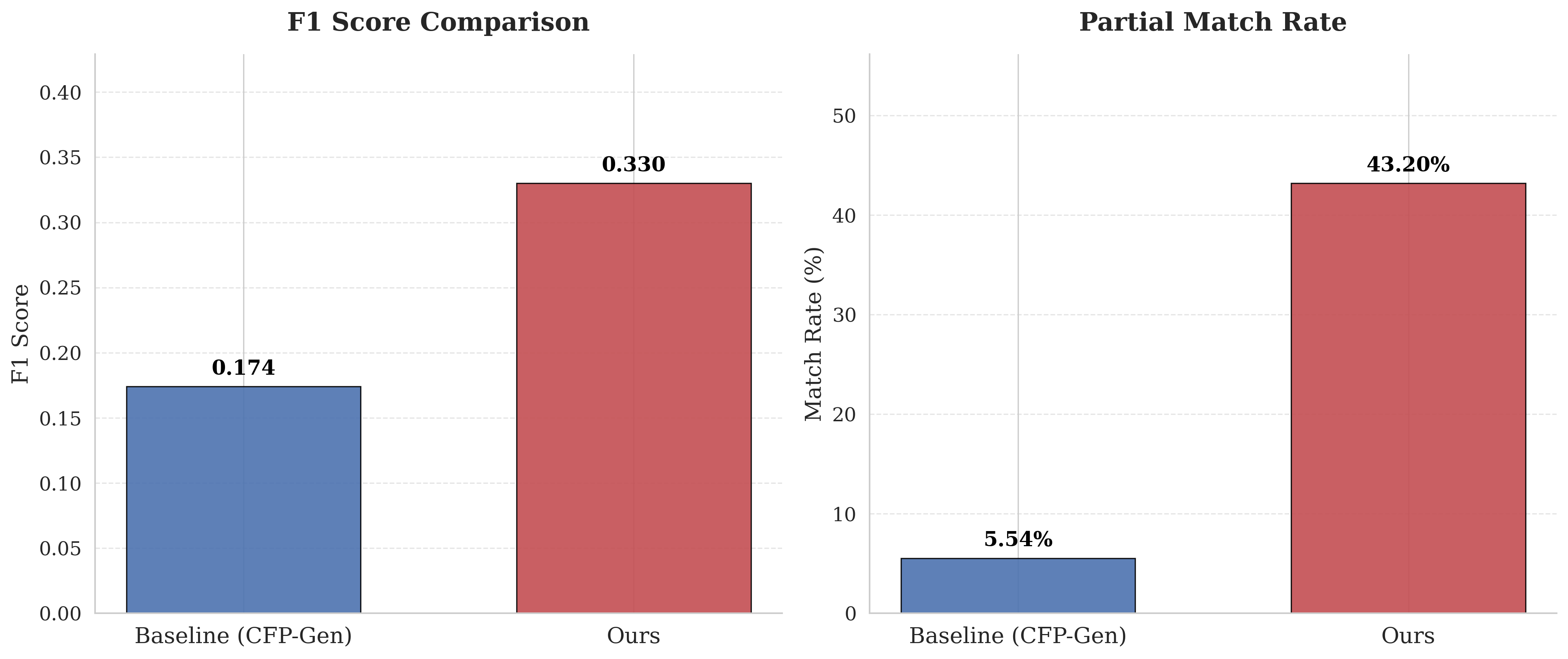}
    \caption{\textbf{Performance on Hypothetical Functional Combinations.} We evaluate the ability to generate proteins for 119 functional combinations not found in nature.}
    \label{fig:hypothetical_perf}
    \vspace{-20pt}
\end{figure}

\textbf{Hypothetical Functional Combinations.}
We further challenge the model to explore previously undefined regions of the functional landscape by generating 10 candidates per combination for 119 synthetically constructed hypothetical combinations (see Appendix \ref{app:test_set_construction}). 
Unfortunately, no generated protein fully satisfies the complete set of constraints, highlighting the inherent difficulty of engineering biologically viable proteins for artificial functional constraints.
Nevertheless, our model exhibits partially correct functional generation even in this severe OOD setting (Fig.\ref{fig:hypothetical_perf}). It significantly outperforms the baseline, raising the F1 score to 0.330 (vs. 0.174) and the Partial Match Rate to 43.20\% (vs. 5.54\%).

\subsection{Case Study}

\begin{figure}[t]
    \centering
    \includegraphics[width=0.9\linewidth]{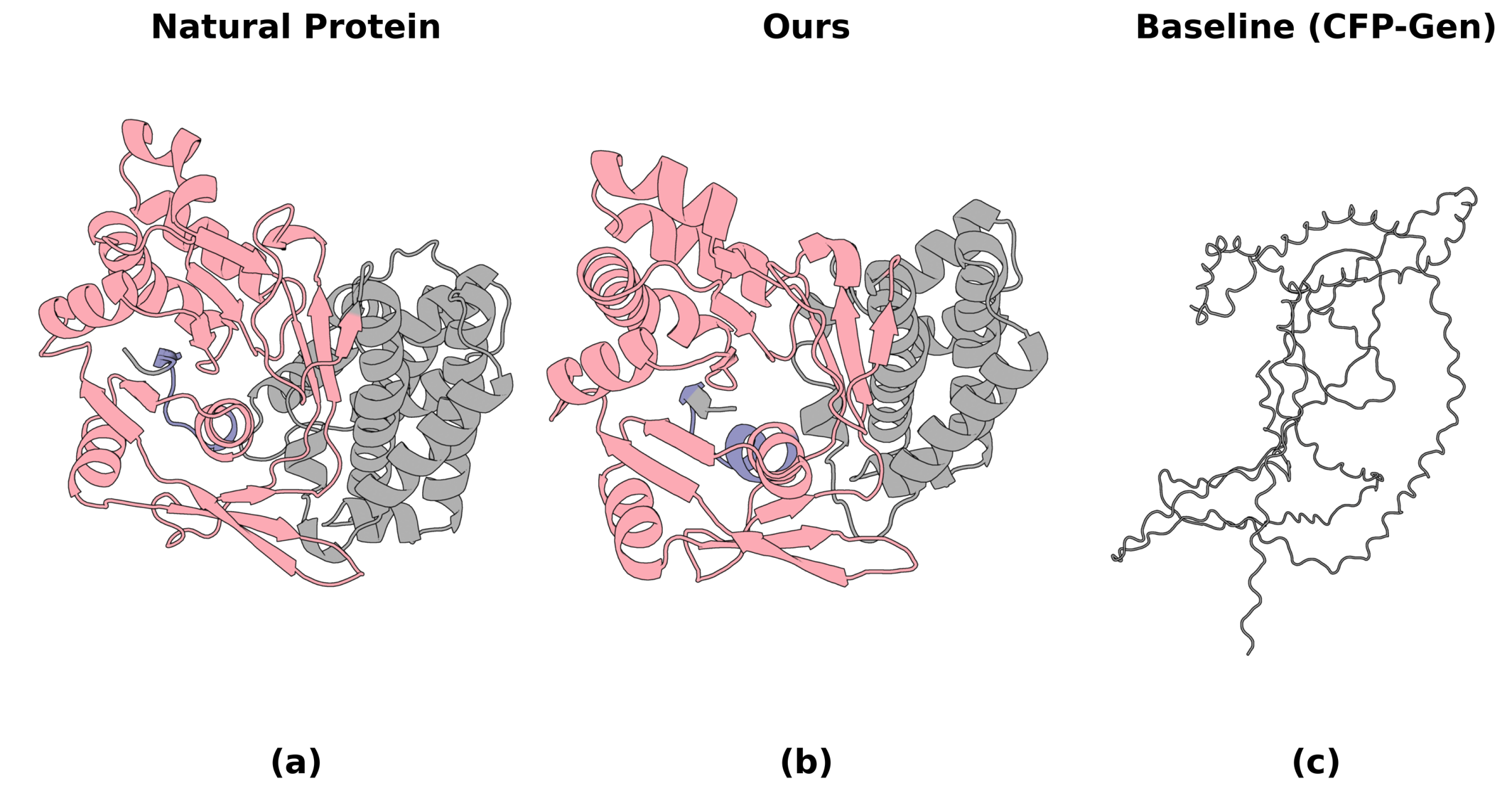}
    \caption{\textbf{Visualization of Multi-Functional Protein Generation for OOD Combinations.} We generate a protein conditioned on the unseen functional combination of \textit{Mannitol-1-phosphate 5-dehydrogenase activity} (GO:0008926) and \textit{NAD binding} (GO:0051287).
    (a) Natural protein structure. (b) Structure generated by our method. (c) Structure generated by the baseline CFP-Gen.
    The \textit{NAD-binding} structural motif is highlighted in red, while the catalytic motif associated with \textit{Mannitol-1-phosphate 5-dehydrogenase activity} is shown in blue.}
    \label{fig:case_study}
    \vspace{-10pt}
\end{figure}

To provide an intuitive illustration of the model’s generative capability, we present a representative case study on protein generation conditioned on an OOD functional combination from the test set.
As illustrated in Fig.~\ref{fig:case_study}, our model successfully generates well-formed local structural motifs that closely resemble the functional motifs observed in natural proteins. In contrast, the baseline fails to fold into a structured protein and collapses into disordered coils lacking defined secondary structure.
Complementing this visual analysis, quantitative metrics further confirm the quality of our generation. The generated protein achieves a pLDDT of 94.9 and a pTM of 0.96, indicating excellent foldability. Crucially, the maximum sequence identity compared to proteins of the same function is merely 32\%. This low homology denotes significant sequence novelty and underscores the model’s capacity for \textit{de novo} functional integration.

\subsection{Ablation Study}

To dissect the contributions of our individual components, we evaluate a two-step version and variants excluding the Functional-Structural Retrieval (FSR) and Local Structure-Function Supervision (LSFS) modules (Table~\ref{tab:ablation}).

\textbf{Efficacy of Co-generation.} The ablation results demonstrate that CodeFP (w/o FSR, LSFS) consistently surpasses the one-step baseline (CFP-Gen) in functionality while exhibiting superior foldability than the two-step version. This validates our hypothesis: explicitly modeling the joint probability offers a superior foundation for functional design compared to one-step generation, while preserving foldability more readily than two-step generation.

\textbf{Dual Role of FSR.} The integration of FSR yields simultaneous gains in functional metrics (F1-Micro: 0.495 vs. 0.465) and foldability (pLDDT $>$ 70: 82.01\% vs. 71.71\%). This dual gain suggests that retrieved structural motifs serve as an essential inductive bias, effectively ground functional semantics, and facilitate the holistic functional co-generation of protein sequence and structure.

\textbf{Distributional Alignment via LSFS.} The deployment of LSFS substantially improves MRR (0.674 vs. 0.534) and reduces MMD (0.101 vs. 0.192). These distributional shifts confirm that the generative distribution aligns more closely with the natural functional protein space, indicating that LSFS provides precise functional supervision that effectively captures the functional semantics inherent in natural proteins.

Our Full Model effectively integrates these mechanisms, yielding the highest functional performance without compromising foldability or distributional fidelity.

\section{Conclusion and Future Work}

In this work, we introduce \textbf{CodeFP}, a novel co-generative PLM framework that unifies sequence and structure generation to advance \textit{de novo} functional protein design.
To extend co-generation to functional design, we propose two critical mechanisms: Functional-Structural Retrieval (FSR), which grounds function semantics by structure motifs, and Local Structure-Function Supervision (LSFS), which mitigates training ambiguity via latent space supervision.
Empirical evaluations on benchmarks demonstrate that CodeFP achieves state-of-the-art performance in both functional consistency and structural foldability, with ablation studies validating the necessity of each component.

While CodeFP advances \textit{de novo} functional protein design, conditioning generation on OOD functional combinations remains challenging. We expect future works on (1) augmenting the current dataset to encompass a broader spectrum of functions, thereby facilitating more rigorous evaluation benchmarks (2) investigating novel function combination design to enhance robustness against OOD shifts (3) applying CodeFP to wet-lab validation empirically substantiate its practical utility and reliability.

\section*{Acknowledgements}
This research is supported by the Innovative Drug Research and Development National Science and Technology Major Project (No. 2025ZD1803101) and PharMolix Inc.

\section*{Impact Statement}
This paper presents work whose goal is to advance the field of Machine Learning. There are many potential societal consequences of our work, none which we feel must be specifically highlighted here.


\bibliography{example_paper}

@article{kortemme2024novo,
  title={De novo protein design—From new structures to programmable functions},
  author={Kortemme, Tanja},
  journal={Cell},
  volume={187},
  number={3},
  pages={526--544},
  year={2024},
  publisher={Elsevier}
}

@article{Progen,
  title={Progen: Language modeling for protein generation},
  author={Madani, Ali and McCann, Bryan and Naik, Nikhil and Keskar, Nitish Shirish and Anand, Namrata and Eguchi, Raphael R and Huang, Po-Ssu and Socher, Richard},
  journal={arXiv preprint arXiv:2004.03497},
  year={2020}
}

@inproceedings{ZymCTRL,
  title={ZymCTRL: a conditional language model for the controllable generation of artificial enzymes},
  author={Munsamy, Geraldene and Lindner, Sebastian and Lorenz, Philipp and Ferruz, Noelia},
  booktitle={NeurIPS machine learning in structural biology workshop},
  year={2022},
  organization={NeurIPS}
}

@article{RFdiffusion,
  title={De novo design of protein structure and function with RFdiffusion},
  author={Watson, Joseph L and Juergens, David and Bennett, Nathaniel R and Trippe, Brian L and Yim, Jason and Eisenach, Helen E and Ahern, Woody and Borst, Andrew J and Ragotte, Robert J and Milles, Lukas F and others},
  journal={Nature},
  volume={620},
  number={7976},
  pages={1089--1100},
  year={2023},
  publisher={Nature Publishing Group UK London}
}

@article{chroma,
  title={Illuminating protein space with a programmable generative model},
  author={Ingraham, John B and Baranov, Max and Costello, Zak and Barber, Karl W and Wang, Wujie and Ismail, Ahmed and Frappier, Vincent and Lord, Dana M and Ng-Thow-Hing, Christopher and Van Vlack, Erik R and others},
  journal={Nature},
  volume={623},
  number={7989},
  pages={1070--1078},
  year={2023},
  publisher={Nature Publishing Group UK London}
}

@article{ProteinMPNN,
  title={Robust deep learning--based protein sequence design using ProteinMPNN},
  author={Dauparas, Justas and Anishchenko, Ivan and Bennett, Nathaniel and Bai, Hua and Ragotte, Robert J and Milles, Lukas F and Wicky, Basile IM and Courbet, Alexis and de Haas, Rob J and Bethel, Neville and others},
  journal={Science},
  volume={378},
  number={6615},
  pages={49--56},
  year={2022},
  publisher={American Association for the Advancement of Science}
}

@article{esm2_esmfold,
  title={Language models of protein sequences at the scale of evolution enable accurate structure prediction},
  author={Lin, Zeming and Akin, Halil and Rao, Roshan and Hie, Brian and Zhu, Zhongkai and Lu, Wenting and Smetanin, Nikita and dos Santos Costa, Allan and Fazel-Zarandi, Maryam and Sercu, Tom and Candido, Sal and others},
  journal={bioRxiv},
  year={2022},
  publisher={Cold Spring Harbor Laboratory}
}

@article{ProtGPT2,
  title={ProtGPT2 is a deep unsupervised language model for protein design},
  author={Ferruz, Noelia and Schmidt, Steffen and H{\"o}cker, Birte},
  journal={Nature communications},
  volume={13},
  number={1},
  pages={4348},
  year={2022},
  publisher={Nature Publishing Group UK London}
}

@article{evodiff,
  title={Protein generation with evolutionary diffusion: sequence is all you need},
  author={Alamdari, Sarah and Thakkar, Nitya and Van Den Berg, Rianne and Tenenholtz, Neil and Strome, Robert and Moses, Alan M and Lu, Alex X and Fusi, Nicol{\`o} and Amini, Ava P and Yang, Kevin K},
  journal={BioRxiv},
  pages={2023--09},
  year={2023},
  publisher={Cold Spring Harbor Laboratory}
}

@article{dplm,
  title={Diffusion language models are versatile protein learners},
  author={Wang, Xinyou and Zheng, Zaixiang and Ye, Fei and Xue, Dongyu and Huang, Shujian and Gu, Quanquan},
  journal={arXiv preprint arXiv:2402.18567},
  year={2024}
}

@article{multiflow,
  title={Generative flows on discrete state-spaces: Enabling multimodal flows with applications to protein co-design},
  author={Campbell, Andrew and Yim, Jason and Barzilay, Regina and Rainforth, Tom and Jaakkola, Tommi},
  journal={arXiv preprint arXiv:2402.04997},
  year={2024}
}

@article{esm3,
  title={Simulating 500 million years of evolution with a language model},
  author={Hayes, Thomas and Rao, Roshan and Akin, Halil and Sofroniew, Nicholas J and Oktay, Deniz and Lin, Zeming and Verkuil, Robert and Tran, Vincent Q and Deaton, Jonathan and Wiggert, Marius and others},
  journal={Science},
  volume={387},
  number={6736},
  pages={850--858},
  year={2025},
  publisher={American Association for the Advancement of Science}
}

@article{dplm2,
  title={Dplm-2: A multimodal diffusion protein language model},
  author={Wang, Xinyou and Zheng, Zaixiang and Ye, Fei and Xue, Dongyu and Huang, Shujian and Gu, Quanquan},
  journal={arXiv preprint arXiv:2410.13782},
  year={2024}
}

@article{proGAN,
  title={Conditional generative modeling for de novo protein design with hierarchical functions},
  author={Kucera, Tim and Togninalli, Matteo and Meng-Papaxanthos, Laetitia},
  journal={Bioinformatics},
  volume={38},
  number={13},
  pages={3454--3461},
  year={2022},
  publisher={Oxford University Press}
}

@inproceedings{cfpgen,
  title={CFP-Gen: Combinatorial Functional Protein Generation via Diffusion Language Models},
  author={Yin, Junbo and Zha, Chao and He, Wenjia and Xu, Chencheng and Gao, Xin},
  booktitle={Forty-second International Conference on Machine Learning},
  year={2025}
}

@article{ProDVa,
  title={Protein design with dynamic protein vocabulary},
  author={Liu, Nuowei and Kuang, Jiahao and Liu, Yanting and Ji, Tao and Sun, Changzhi and Lan, Man and Wu, Yuanbin},
  journal={arXiv preprint arXiv:2505.18966},
  year={2025}
}

@article{pinal,
  title={Toward de novo protein design from natural language},
  author={Dai, Fengyuan and You, Shiyang and Zhu, Yudian and Gao, Yuan and Fu, Lihao and Zhou, Xibin and Su, Jin and Wang, Chentong and Fan, Yuliang and Ma, Xiaoxiao and others},
  journal={BioRxiv},
  pages={2024--08},
  year={2024},
  publisher={Cold Spring Harbor Laboratory}
}

@article{lfq,
  title={Language Model Beats Diffusion--Tokenizer is Key to Visual Generation},
  author={Yu, Lijun and Lezama, Jos{\'e} and Gundavarapu, Nitesh B and Versari, Luca and Sohn, Kihyuk and Minnen, David and Cheng, Yong and Birodkar, Vighnesh and Gupta, Agrim and Gu, Xiuye and others},
  journal={arXiv preprint arXiv:2310.05737},
  year={2023}
}

@article{interproscan,
  title={InterProScan 5: genome-scale protein function classification},
  author={Jones, Philip and Binns, David and Chang, Hsin-Yu and Fraser, Matthew and Li, Weizhong and McAnulla, Craig and McWilliam, Hamish and Maslen, John and Mitchell, Alex and Nuka, Gift and others},
  journal={Bioinformatics},
  volume={30},
  number={9},
  pages={1236--1240},
  year={2014},
  publisher={Oxford University Press}
}

@article{pdb,
  title={The protein data bank},
  author={Berman, Helen M and Westbrook, John and Feng, Zukang and Gilliland, Gary and Bhat, Talapady N and Weissig, Helge and Shindyalov, Ilya N and Bourne, Philip E},
  journal={Nucleic acids research},
  volume={28},
  number={1},
  pages={235--242},
  year={2000},
  publisher={Oxford University Press}
}

@article{afdb,
  title={AlphaFold Protein Structure Database in 2024: providing structure coverage for over 214 million protein sequences},
  author={Varadi, Mihaly and Bertoni, Damian and Magana, Paulyna and Paramval, Urmila and Pidruchna, Ivanna and Radhakrishnan, Malarvizhi and Tsenkov, Maxim and Nair, Sreenath and Mirdita, Milot and Yeo, Jingi and others},
  journal={Nucleic acids research},
  volume={52},
  number={D1},
  pages={D368--D375},
  year={2024},
  publisher={Oxford University Press}
}

@article{deepgo,
  title={Deepgo-se: Protein function prediction as approximate semantic entailment},
  author={Kulmanov, Maxat and Guzm{\'a}n-Vega, Francisco J and Roggli, Paula Duek and Lane, Lydie and Arold, Stefan T and Hoehndorf, Robert},
  journal={bioRxiv},
  pages={2023--09},
  year={2023},
  publisher={Cold Spring Harbor Laboratory}
}

@article{mmseqs2,
  title={MMseqs2 enables sensitive protein sequence searching for the analysis of massive data sets},
  author={Steinegger, Martin and S{\"o}ding, Johannes},
  journal={Nature biotechnology},
  volume={35},
  number={11},
  pages={1026--1028},
  year={2017},
  publisher={Nature Publishing Group US New York}
}

@article{l21,
  title={Design of a potent interleukin-21 mimic for cancer immunotherapy},
  author={Chun, Jung-Ho and Lim, Birkley S and Roy, Suyasha and Walsh, Michael J and Abhiraman, Gita C and Zhangxu, Kevin and Atajanova, Tavus and Revach, Or-Yam and Clark, Elisa C and Li, Peng and others},
  journal={Science immunology},
  volume={10},
  number={111},
  pages={eadx1582},
  year={2025},
  publisher={American Association for the Advancement of Science}
}

@article{de1,
  title={Biocatalytic asymmetric synthesis of chiral amines from ketones applied to sitagliptin manufacture},
  author={Savile, Christopher K and Janey, Jacob M and Mundorff, Emily C and Moore, Jeffrey C and Tam, Sarena and Jarvis, William R and Colbeck, Jeffrey C and Krebber, Anke and Fleitz, Fred J and Brands, Jos and others},
  journal={Science},
  volume={329},
  number={5989},
  pages={305--309},
  year={2010},
  publisher={American Association for the Advancement of Science}
}

@article{scaffolding1,
  title={Scaffolding protein functional sites using deep learning},
  author={Wang, Jue and Lisanza, Sidney and Juergens, David and Tischer, Doug and Watson, Joseph L and Castro, Karla M and Ragotte, Robert and Saragovi, Amijai and Milles, Lukas F and Baek, Minkyung and others},
  journal={Science},
  volume={377},
  number={6604},
  pages={387--394},
  year={2022},
  publisher={American Association for the Advancement of Science}
}

@article{flow4backbone,
  title={Se (3)-stochastic flow matching for protein backbone generation},
  author={Bose, Avishek Joey and Akhound-Sadegh, Tara and Huguet, Guillaume and Fatras, Kilian and Rector-Brooks, Jarrid and Liu, Cheng-Hao and Nica, Andrei Cristian and Korablyov, Maksym and Bronstein, Michael and Tong, Alexander},
  journal={arXiv preprint arXiv:2310.02391},
  year={2023}
}

@inproceedings{SLM,
  title={Structure Language Models for Protein Conformation Generation},
  author={Lu, Jiarui and Chen, Xiaoyin and Lu, Stephen Zhewen and Shi, Chence and Guo, Hongyu and Bengio, Yoshua and Tang, Jian},
  booktitle={The Thirteenth International Conference on Learning Representations},
  year={2024}
}

@article{co_gen_go,
  title={Generating functional and multistate proteins with a multimodal diffusion transformer},
  author={Jing, Bowen and Sappington, Anna and Bafna, Mihir and Shah, Ravi and Tang, Adrina and Krishna, Rohith and Klivans, Adam and Diaz, Daniel J and Berger, Bonnie},
  journal={bioRxiv},
  year={2025}
}

@article{enzyme_app_1,
  title={Characterization and engineering of a plastic-degrading aromatic polyesterase},
  author={Austin, Harry P and Allen, Mark D and Donohoe, Bryon S and Rorrer, Nicholas A and Kearns, Fiona L and Silveira, Rodrigo L and Pollard, Benjamin C and Dominick, Graham and Duman, Ramona and El Omari, Kamel and others},
  journal={Proceedings of the National Academy of Sciences},
  volume={115},
  number={19},
  pages={E4350--E4357},
  year={2018},
  publisher={National Academy of Sciences}
}

@article{enzyme_app_2,
  title={Automated design of efficient and functionally diverse enzyme repertoires},
  author={Khersonsky, Olga and Lipsh, Rosalie and Avizemer, Ziv and Ashani, Yacov and Goldsmith, Moshe and Leader, Haim and Dym, Orly and Rogotner, Shelly and Trudeau, Devin L and Prilusky, Jaime and others},
  journal={Molecular cell},
  volume={72},
  number={1},
  pages={178--186},
  year={2018},
  publisher={Elsevier}
}

@article{therapeutic_app1,
  title={Rational design and engineering of therapeutic proteins},
  author={Marshall, Shannon A and Lazar, Greg A and Chirino, Arthur J and Desjarlais, John R},
  journal={Drug discovery today},
  volume={8},
  number={5},
  pages={212--221},
  year={2003},
  publisher={Elsevier}
}

@article{antibody_app1,
  title={Computationally designed bispecific antibodies using negative state repertoires},
  author={Leaver-Fay, Andrew and Froning, Karen J and Atwell, Shane and Aldaz, Hector and Pustilnik, Anna and Lu, Frances and Huang, Flora and Yuan, Richard and Hassanali, Saleema and Chamberlain, Aaron K and others},
  journal={Structure},
  volume={24},
  number={4},
  pages={641--651},
  year={2016},
  publisher={Elsevier}
}

@article{dl1,
  title={Rapid evolution of a protein in vitro by DNA shuffling},
  author={Stemmer, Willem PC},
  journal={Nature},
  volume={370},
  number={6488},
  pages={389--391},
  year={1994},
  publisher={Nature Publishing Group UK London}
}

@article{dl2,
  title={Machine-learning-guided directed evolution for protein engineering},
  author={Yang, Kevin K and Wu, Zachary and Arnold, Frances H},
  journal={Nature methods},
  volume={16},
  number={8},
  pages={687--694},
  year={2019},
  publisher={Nature Publishing Group US New York}
}

@article{denovo_1,
  title={De novo design of luciferases using deep learning},
  author={Yeh, Andy Hsien-Wei and Norn, Christoffer and Kipnis, Yakov and Tischer, Doug and Pellock, Samuel J and Evans, Declan and Ma, Pengchen and Lee, Gyu Rie and Zhang, Jason Z and Anishchenko, Ivan and others},
  journal={Nature},
  volume={614},
  number={7949},
  pages={774--780},
  year={2023},
  publisher={Nature Publishing Group UK London}
}

@article{go,
  title={Gene ontology: tool for the unification of biology},
  author={Ashburner, Michael and Ball, Catherine A and Blake, Judith A and Botstein, David and Butler, Heather and Cherry, J Michael and Davis, Allan P and Dolinski, Kara and Dwight, Selina S and Eppig, Janan T and others},
  journal={Nature genetics},
  volume={25},
  number={1},
  pages={25--29},
  year={2000},
  publisher={Nature Publishing Group}
}

@article{discrete_diffusion,
  title={Structured denoising diffusion models in discrete state-spaces},
  author={Austin, Jacob and Johnson, Daniel D and Ho, Jonathan and Tarlow, Daniel and Van Den Berg, Rianne},
  journal={Advances in neural information processing systems},
  volume={34},
  pages={17981--17993},
  year={2021}
}

@article{uniprot2025uniprot,
  title={UniProt: the universal protein knowledgebase in 2025},
  journal={Nucleic acids research},
  volume={53},
  number={D1},
  pages={D609--D617},
  year={2025},
  publisher={Oxford University Press}
}

@article{blum2025interpro,
  title={InterPro: the protein sequence classification resource in 2025<? mode longmeta?>},
  author={Blum, Matthias and Andreeva, Antonina and Florentino, Laise Cavalcanti and Chuguransky, Sara Rocio and Grego, Tiago and Hobbs, Emma and Pinto, Beatriz Lazaro and Orr, Ailsa and Paysan-Lafosse, Typhaine and Ponamareva, Irina and others},
  journal={Nucleic acids research},
  volume={53},
  number={D1},
  pages={D444--D456},
  year={2025},
  publisher={Oxford University Press}
}

@article{coflow,
  title={Co-design protein sequence and structure in discrete space via generative flow},
  author={Yang, Sen and Ju, Lingli and Cheng, Peng and Zhou, JiangLin and Cai, Yamin and Feng, Dawei},
  journal={Bioinformatics},
  volume={41},
  number={5},
  pages={btaf248},
  year={2025},
  publisher={Oxford University Press}
}

@article{dl_r1,
  title={Machine-learning-guided directed evolution for protein engineering},
  author={Yang, Kevin K and Wu, Zachary and Arnold, Frances H},
  journal={Nature methods},
  volume={16},
  number={8},
  pages={687--694},
  year={2019},
  publisher={Nature Publishing Group US New York}
}

@article{prollama,
  title={Prollama: A protein large language model for multi-task protein language processing},
  author={Lv, Liuzhenghao and Lin, Zongying and Li, Hao and Liu, Yuyang and Cui, Jiaxi and Chen, Calvin Yu-Chian and Yuan, Li and Tian, Yonghong},
  journal={IEEE Transactions on Artificial Intelligence},
  year={2025},
  publisher={IEEE}
}
\bibliographystyle{icml2026}

\newpage
\appendix
\onecolumn

\section{Evaluation Metrics}
\label{app:metrics}

We comprehensively evaluate the generated proteins across three dimensions: Sequence Plausibility (distributional similarity to natural proteins), Functional Consistency (alignment with target functional constraints), and Structural Realizability (physical foldability).

\subsection{Sequence Distribution Metrics}

To quantify how well the generated proteins capture the biophysical properties of natural proteins, we measure the distributional discrepancy between the full set of generated sequences $\mathcal{G}$ and natural sequences $\mathcal{P}$. We utilize Maximum Mean Discrepancy (MMD) with sequence embeddings derived from normalized Spectrum Mapping (k-mer frequencies). We report MMD with two kernels:
\begin{itemize}
    \item \textbf{Linear MMD ($\text{MMD}_{\text{lin}}$):} Measures the Euclidean distance between the mean embeddings of the real and generated distributions:\begin{equation}\text{MMD}_{\text{lin}}(\mathcal{P}, \mathcal{G}) = |\mu_\mathcal{P} - \mu_\mathcal{G}|2\end{equation}where $\mu_\mathcal{P}$ and $\mu_\mathcal{G}$ are the means of the sequence embeddings.
    \item \textbf{Gaussian MMD ($\text{MMD}_{\text{rbf}}$):} Incorporates a Radial Basis Function (RBF) kernel $k(x, y) = \exp(-\gamma \|x - y\|^2)$ to capture higher-order distributional moments. The bandwidth $\gamma$ is determined via the median heuristic. 
\end{itemize}
Lower MMD values indicate that the generated sequences share similar statistical properties with natural proteins.

\subsection{Functional Consistency Metrics}

To assess whether the generated sequences satisfy the specified functional conditions, we employ two complementary evaluation strategies: oracle-based classification and distribution-based ranking.

\textbf{Oracle-based Metrics.} We utilize a pre-trained state-of-the-art function prediction model as an oracle to classify the generated sequences. By comparing the predicted labels against the input conditional labels, we report the following standard metrics:
\begin{itemize}
    \item \textbf{Macro/Micro F1-score:} To balance performance across classes with varying frequencies, we report both Macro-F1 (arithmetic mean of per-class F1) and Micro-F1 (global calculation based on total true/false positives/negatives).
    
    \item \textbf{Macro AUPR \& AUC:} We compute the Area Under the Precision-Recall Curve (AUPR) and the Receiver Operating Characteristic Curve (AUC), averaged across all classes.
\end{itemize}

\textbf{Distribution-based Metric.} We evaluate the distributional alignment between generated and natural sequences within the same functional category. We utilize a \textbf{Mean Reciprocal Rank (MRR)} metric based on the Maximum Mean Discrepancy (MMD). Let $\mathcal{P}_c$ and $\mathcal{G}_c$ denote the sets of real and generated sequences, respectively, for a specific function label $c \in \{1, \dots, C\}$. We compute the linear MMD distance between the real set of class $c$ ($\mathcal{P}_c$) and the generated sets of all classes ($\mathcal{G}_{c'}$ for all $c'$). If the generation is distinct and accurate, $\mathcal{G}_c$ should be closest to $\mathcal{P}_c$. The MRR is defined as:\begin{equation}\text{MRR}(\mathcal{G}, \mathcal{P}) = \frac{1}{C} \sum_{c=1}^{C} \frac{1}{\text{rank}_{\mathcal{G}}(\text{MMD}(\mathcal{G}c, \mathcal{P}c))}\end{equation}where $\text{rank}_{\mathcal{G}}(\cdot)$ is the rank of the distance $\text{MMD}(\mathcal{G}_c, \mathcal{P}_c)$ among the set of distances $\left\{ \text{MMD}(\mathcal{G}_{c'}, \mathcal{P}_c) \right\}_{c'=1}^C$. An MRR of 1.0 indicates perfect functional mode matching.

\subsection{Structural Realizability Metrics}

Since the generated outputs are primary sequences, we assess their foldability by predicting their 3D structures using ESMFold. We utilize two confidence metrics provided by the folding engine:
\begin{itemize}
    \item \textbf{pLDDT:} The predicted Local Distance Difference Test score. We calculate the mean pLDDT per protein. A score $>70$ indicates a high-confidence prediction, suggesting the sequence adopts a stable local structure.
    
    \item \textbf{pTM:} The predicted Template Modeling score, which estimates the global topological accuracy. We consider sequences with $\text{pTM} > 0.5$ as having a likely correct global fold. 
\end{itemize}

\section{Extensive Analysis}

To further investigate the relationship between model generation capabilities and the characteristics of functional labels, we conducted a comprehensive analysis based on the semantic properties of the GO and the distributional properties of the training data.

\subsection{Performance vs. Semantic Difficulty and Oracle Bias}
\label{app:semantic_difficulty}

We first analyze how model performance (F1-score and Recall) varies with the Semantic Difficulty of the input condition. We define the input's semantic difficulty as the mean semantic distance of the requested GO label combination. The semantic distance between two GO labels is calculated as the shortest path length on the GO Directed Acyclic Graph (DAG), where edges represent "is-a" relationships. For a set of input labels, the mean distance is the average of pairwise distances between all labels in the set.

\begin{figure}[h] \centering \begin{subfigure}{0.48\textwidth} \includegraphics[width=\linewidth]{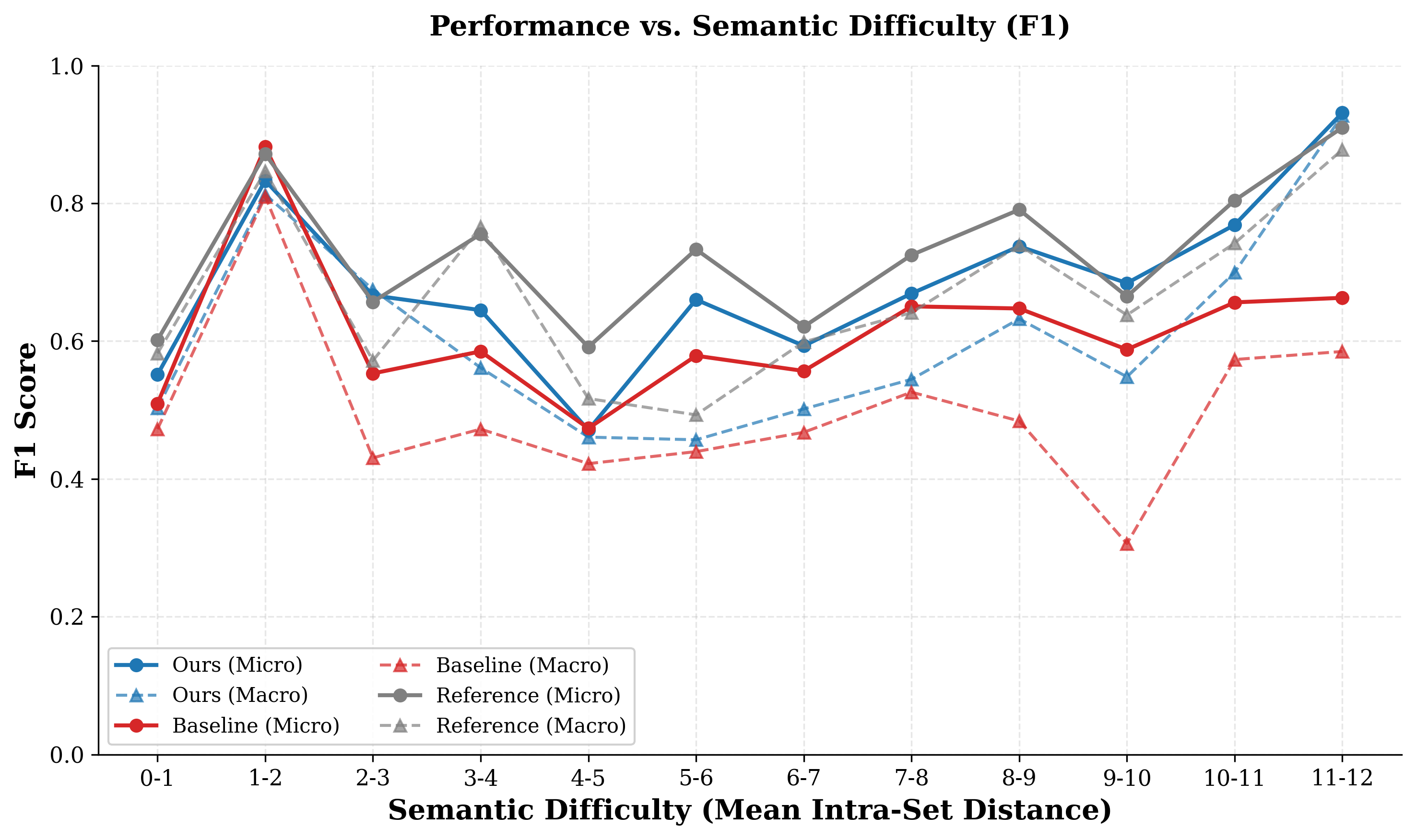} \caption{F1 Score vs. Semantic Difficulty} \end{subfigure} \hfill \begin{subfigure}{0.48\textwidth} \includegraphics[width=\linewidth]{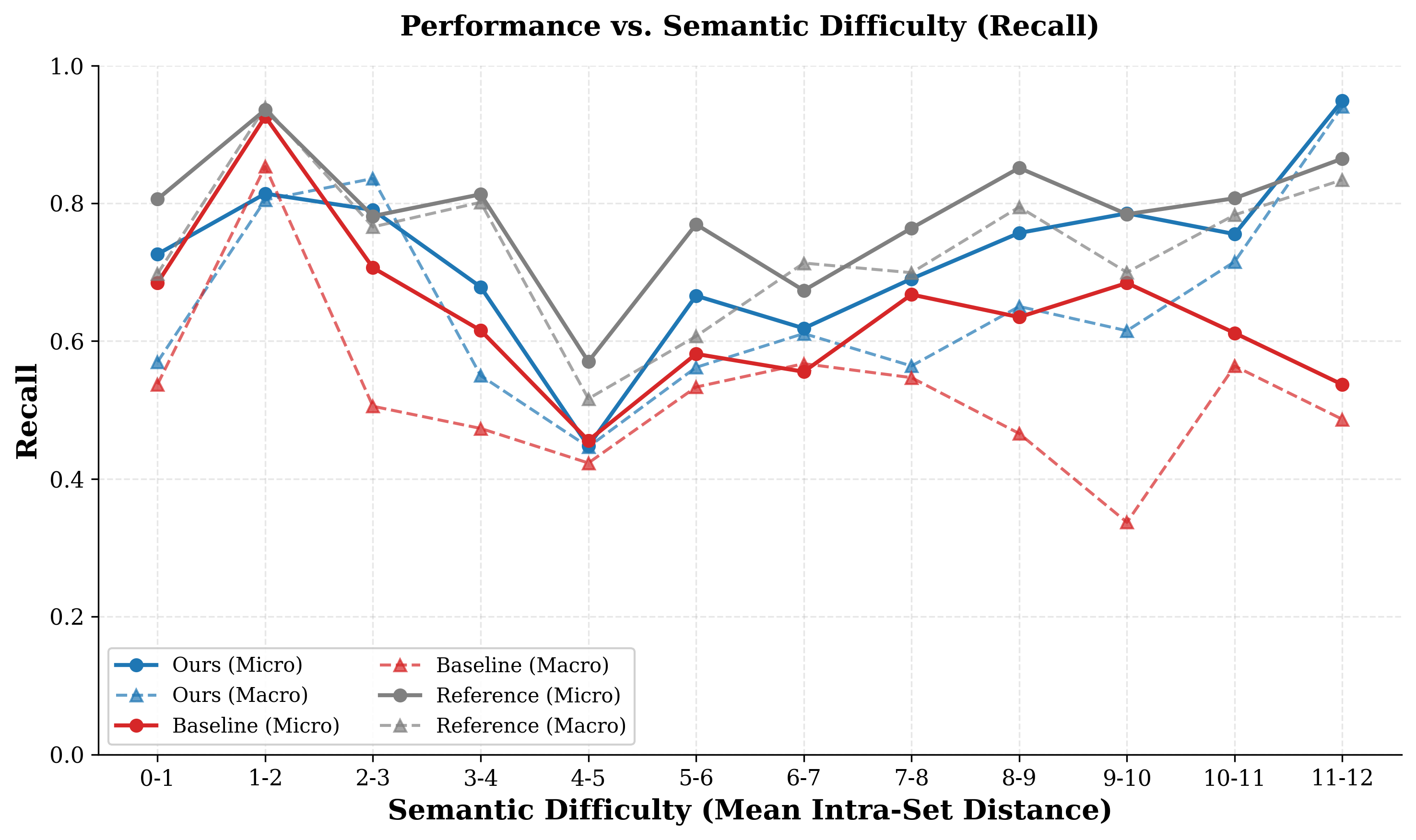} \caption{Recall vs. Semantic Difficulty} \end{subfigure} \caption{\textbf{Performance variation across Semantic Difficulty.} The x-axis represents the mean intra-set semantic distance of input GO labels. We observe a strong correlation between the generative models' performance and the Reference (Oracle) model's performance.} \label{fig:perf_vs_difficulty} \end{figure}

As shown in Fig.\ref{fig:perf_vs_difficulty}, the fluctuation in F1-score and Recall for both our model and the baseline (CFP-Gen) does not strictly correlate with increased difficulty. Instead, it exhibits a strong correlation with the performance of the Reference (Oracle) model. Notably, in the region where the mean semantic distance is 4–5, both the baseline and our model suffer a significant performance drop. This decline coincides with a sharp drop in the Reference model's performance. This suggests that the current evaluation metrics are potentially bottlenecked by the capability of the functional predictor (Oracle), limiting the assessed performance of generative models in specific semantic regions.

\subsection{Analysis of Performance Gap Across Distributional and Biological Properties}
Given the potential bias in absolute evaluation metrics identified above, we further analyze the \textbf{Performance Gap ($\Delta$)}, defined as the score difference between the generative model and the Reference. We compare our model against the baseline across five metrics spanning data distribution and biological significance:\begin{enumerate}\item \textbf{Co-occurrence Strength (Typicality):} Measures how often label pairs appear together. For a pair of labels, it is calculated as their intersection count in the training set divided by the sum of their individual counts.\item \textbf{Train Frequency:} The $\log_{10}$ of the total occurrence count of the label in the training set.\item \textbf{Specificity (IDF):} The Inverse Document Frequency, treating GO labels as words and proteins as documents, calculated as $\log_{10}(N / \text{count})$, where $N$ is the total number of training samples.\item \textbf{Semantic Difficulty (Avg Distance):} As defined in \ref{app:semantic_difficulty}.\item \textbf{Annotation Specificity (Avg Depth):} The average depth of the labels in the GO hierarchy, defined as the shortest path distance from the root node (GO:0003674).\end{enumerate}\begin{figure}[t!]\centering\begin{subfigure}{0.32\textwidth}\includegraphics[width=\linewidth]{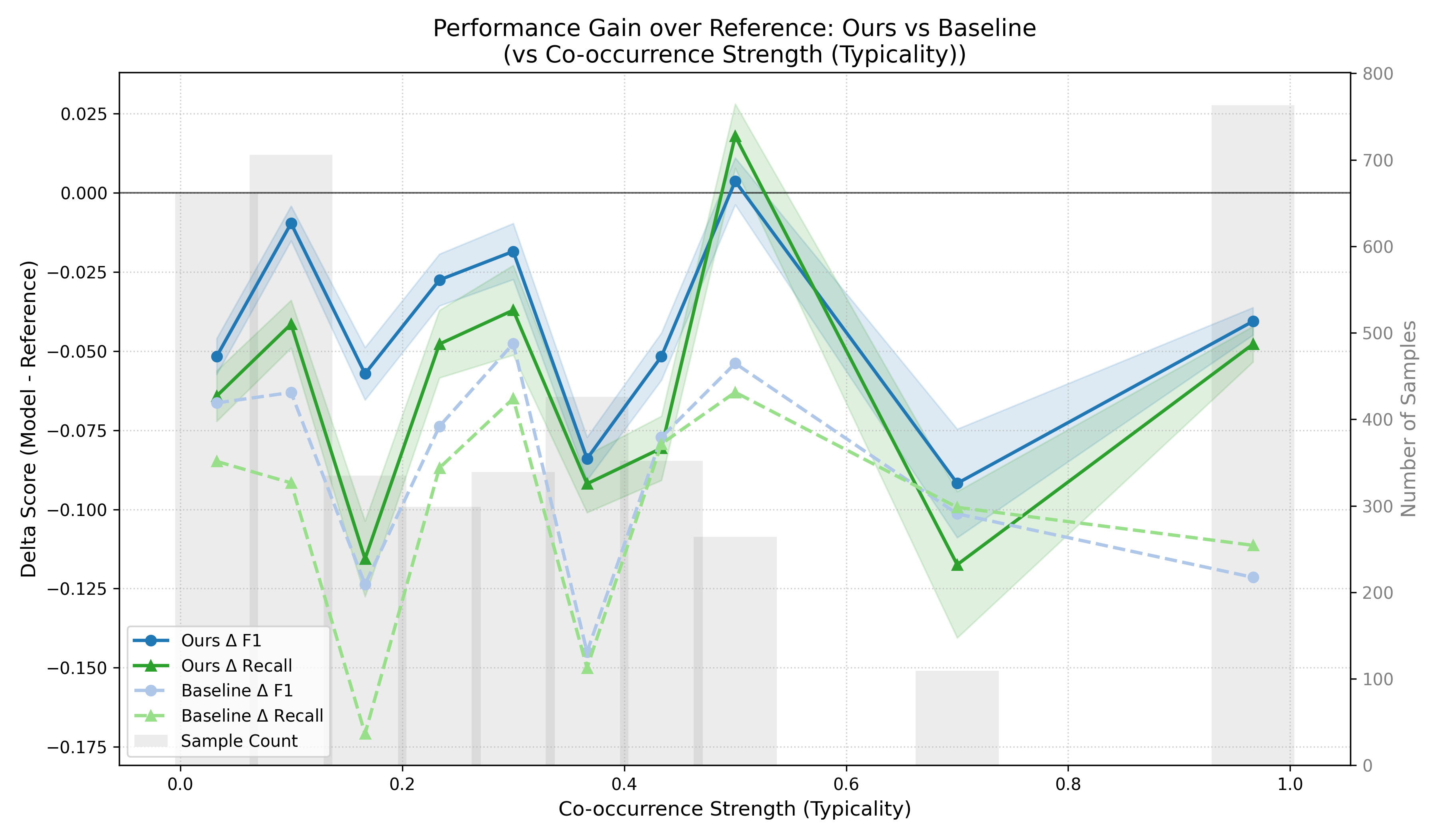}\caption{vs. Co-occurrence Strength}\end{subfigure}\hfill\begin{subfigure}{0.32\textwidth}\includegraphics[width=\linewidth]{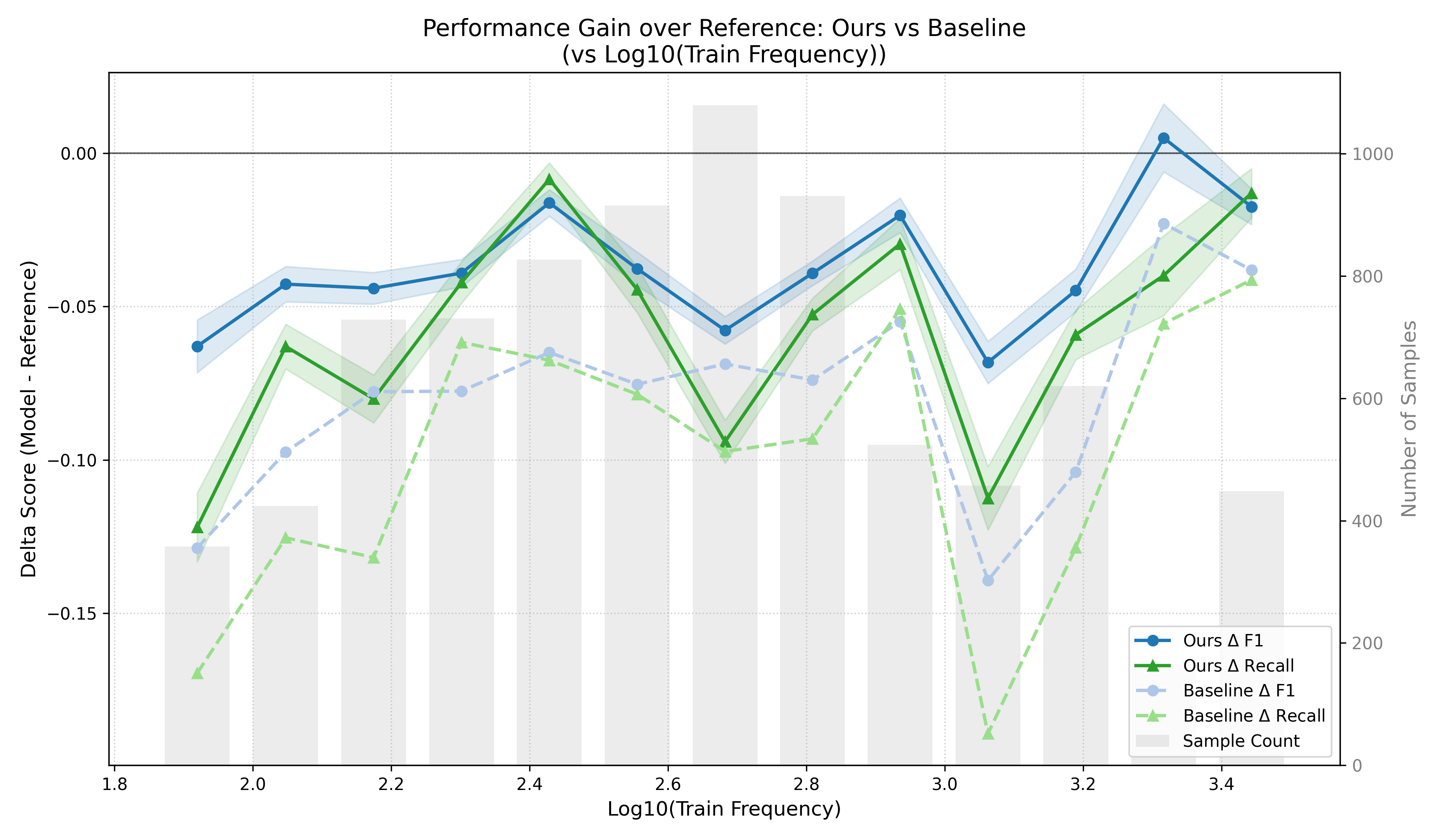}\caption{vs. Train Frequency ($\log_{10}$)}\end{subfigure}\hfill\begin{subfigure}{0.32\textwidth}\includegraphics[width=\linewidth]{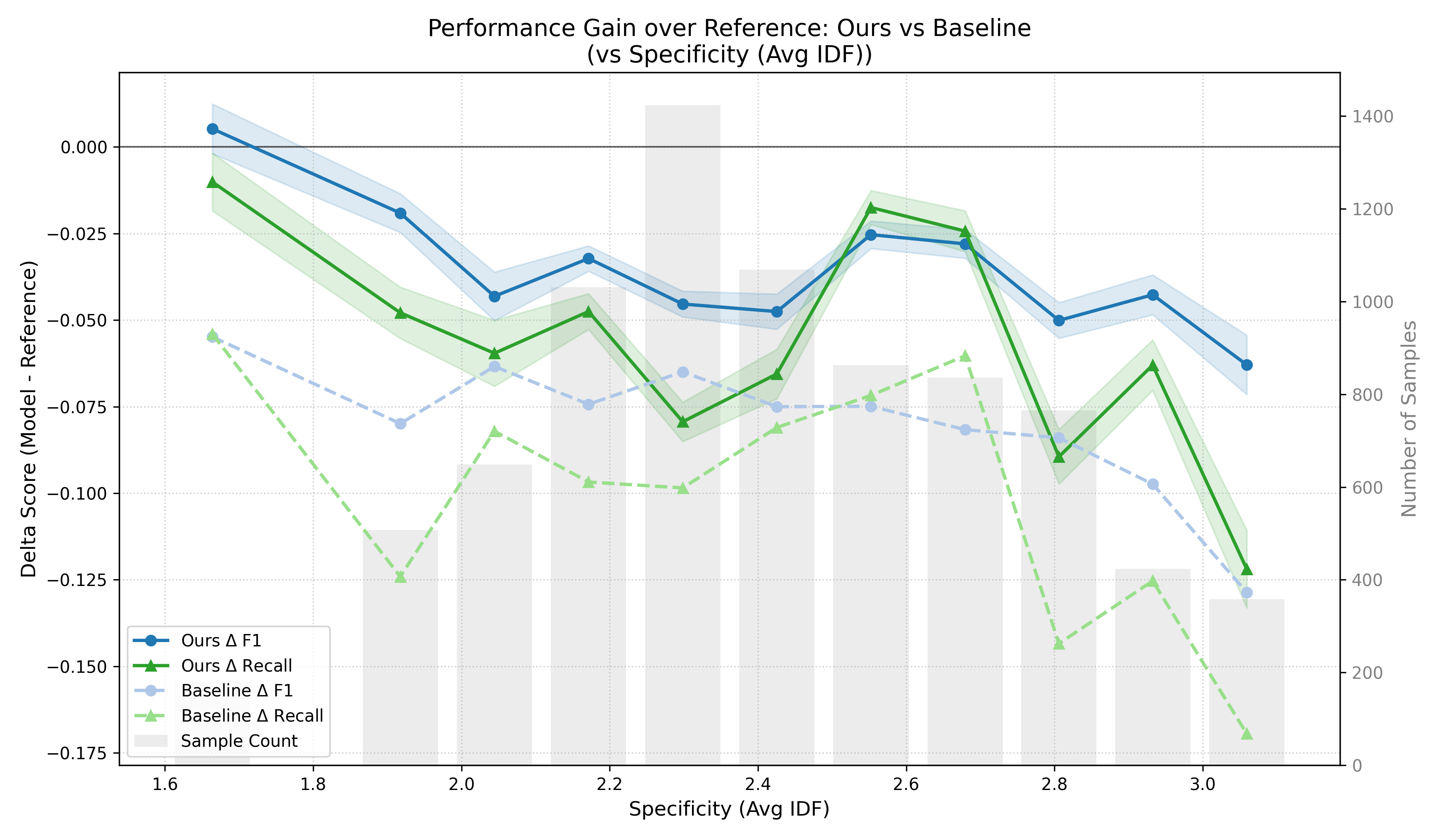}\caption{vs. Specificity (IDF)}\end{subfigure}\vspace{1em}
\begin{subfigure}{0.48\textwidth}
    \includegraphics[width=\linewidth]{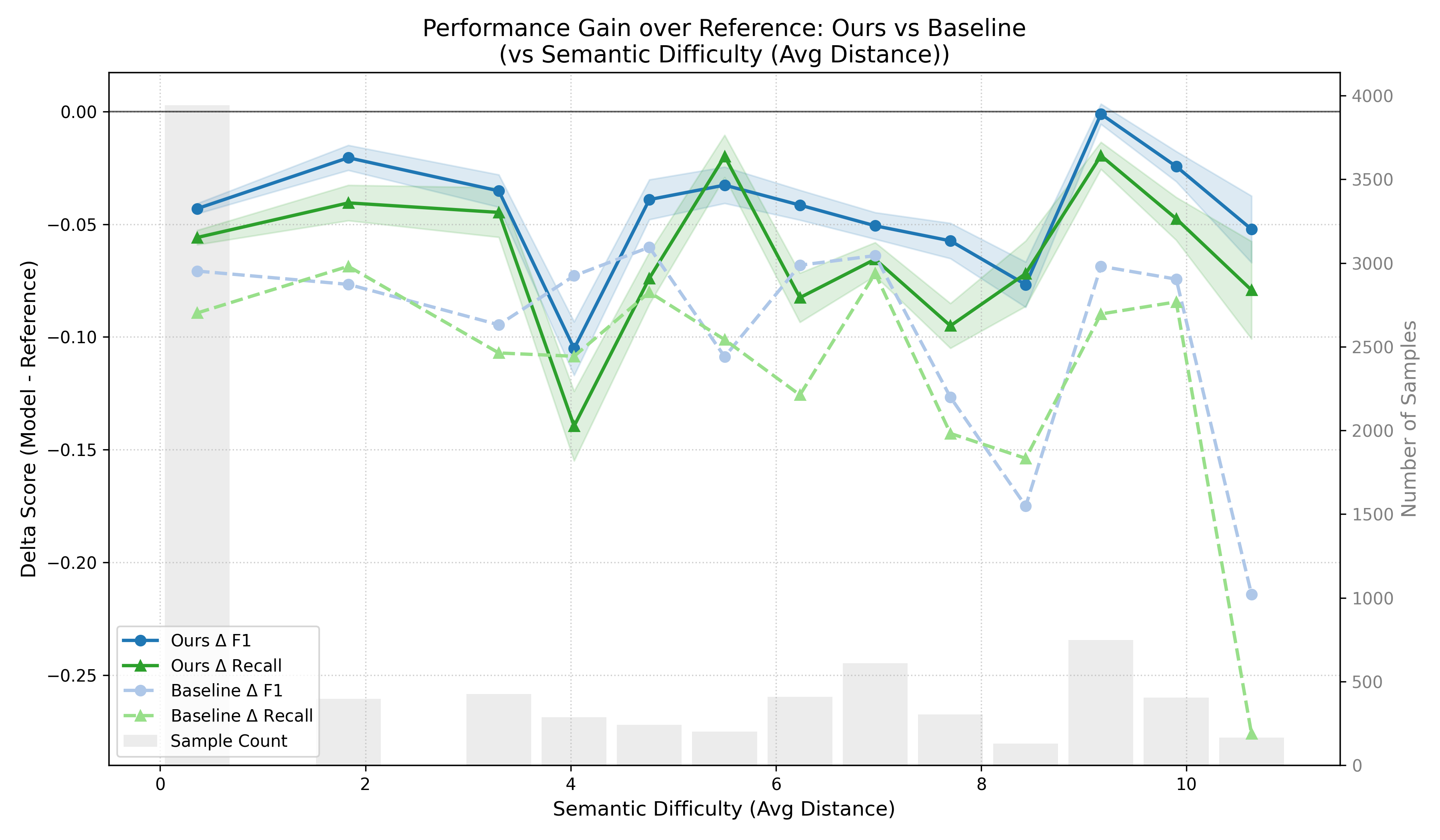}
    \caption{vs. Semantic Difficulty (Distance)}
\end{subfigure}
\hfill
\begin{subfigure}{0.48\textwidth}
    \includegraphics[width=\linewidth]{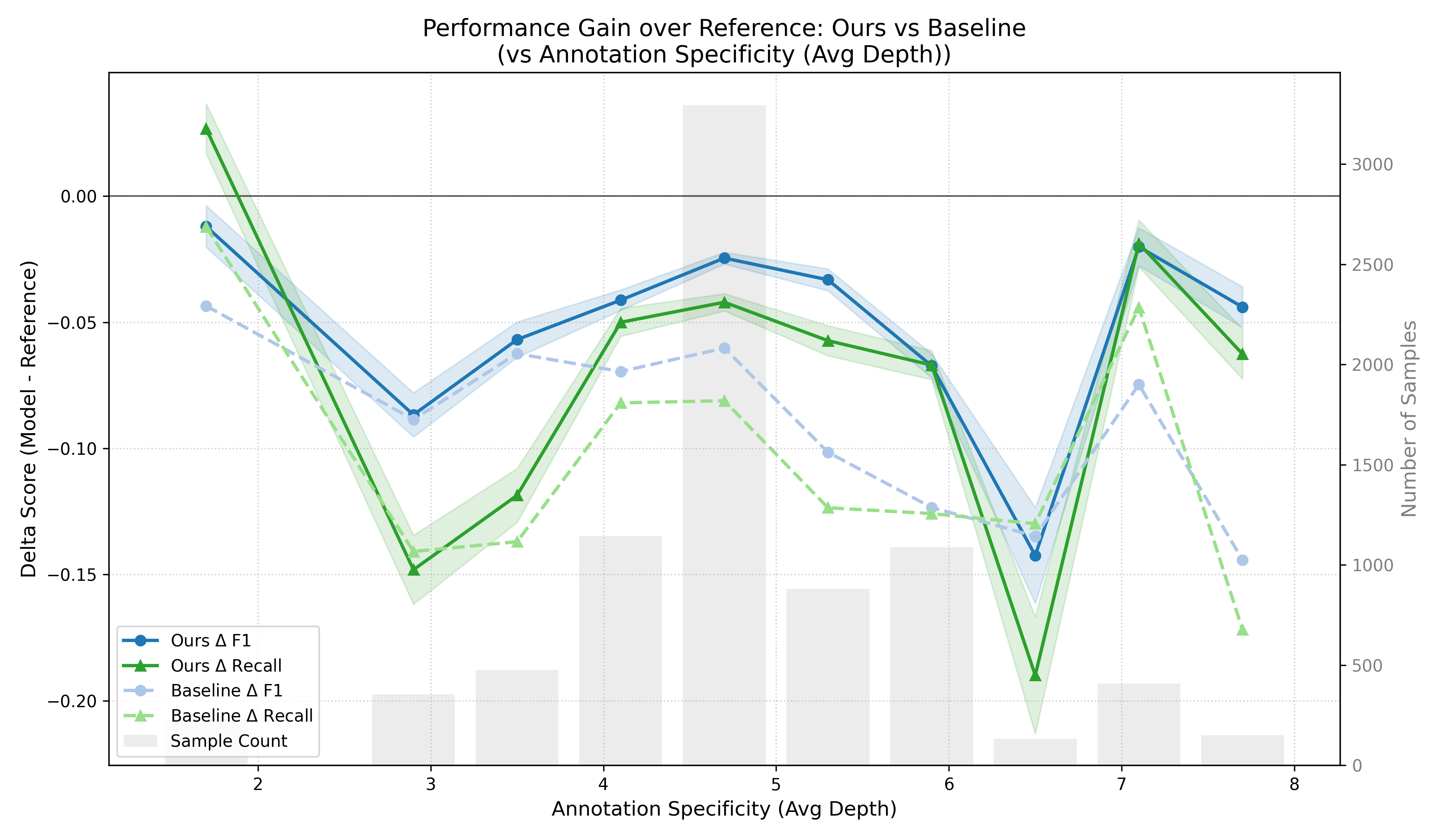}
    \caption{vs. Annotation Specificity (Depth)}
\end{subfigure}

\caption{\textbf{Delta Performance ($\Delta$F1 and $\Delta$Recall) across different functional properties.} The curves represent the performance gap relative to the Reference. Our model (solid lines) consistently achieves a smaller gap (higher values) compared to the baseline (dashed lines).}
\label{fig:delta_analysis}
\end{figure}
As illustrated in Fig.\ref{fig:delta_analysis}, our model consistently exhibits higher $\Delta$ values than the baseline across the majority of metric intervals. The performance trends of our model generally mirror those of the baseline, particularly in Co-occurrence Strength and Depth.

\section{Construction of the Hypothetical Function Combination Dataset} \label{app:test_set_construction}

\textbf{Candidate Selection via Structural Conservation.} We curate a pool of 21 GO terms exhibiting high motif structural stability, specifically selecting those with a median Root Mean Square Deviation (RMSD) below 0.5 \AA\ in the training set. This criterion ensures that the selected functions correspond to highly conserved local structures. From this pool, we generate pairwise combinations through random sampling, filtering out any pairs that co-occur in the training distribution. This procedure yields a final test set of 119 hypothetical GO label combinations.

\textbf{Dataset Statistics.} The resulting dataset exhibits a mean semantic distance of 8.7 and a mean semantic depth of 5.1. These statistics indicate that the selected combinations possess high functional specificity (high depth) while maintaining weak functional correlation (large semantic distance). 

\section{Hyperparameter Details}
\label{app:hypermeter}

\textbf{Model Architecture and Training.} CodeFP is initialized with the DPLM-2 (650M) foundation, scaling to a total of 1.59B parameters. The training process was executed on a cluster of 4 NVIDIA A800 GPUs for approximately 60 hours. Optimization was performed using AdamW with a peak learning rate of $4 \times 10^{-5}$ and a global batch size of 2,048 tokens; all other hyperparameters remain consistent with the base model configuration.

\textbf{Impact of LSFS Configuration.} We investigate the sensitivity of Local Structure-Function Supervision (LSFS) to the weighting coefficient $\gamma$ and class-frequency weighting (Table \ref{tab:lsfs_hyperparam}). A key finding is that unweighted LSFS suffers from poor tail-class performance; applying frequency-based weighting ($\gamma=1.0$) substantially improves F1-Macro from 0.374 to 0.419, effectively mitigating optimization bias in the long-tailed functional distribution. Furthermore, increasing $\gamma$ to 2.0 yields consistent gains in both metrics, suggesting that LSFS offers an accurate supervision signal that guides functional alignment.

\begin{table}[t]
\centering
\caption{\textbf{Impact of LSFS Hyperparameters.} We evaluate the sensitivity of functional alignment (F1-Micro and F1-Macro) to the LSFS loss weighting coefficient ($\gamma$) and the class-frequency weighting strategy.}

\label{tab:lsfs_hyperparam}
\begin{small}
\begin{sc}
\begin{tabular}{lccc}
\toprule
\textbf{Configuration} & \textbf{$\gamma$ (LSFS)} & \textbf{F1-Micro} & \textbf{F1-Macro} \\
\midrule
LSFS (Unweighted) & 1.0 & 0.465 & 0.374 \\
LSFS (Weighted) & 1.0 & 0.474 & 0.419 \\
LSFS (Weighted) & 2.0 & 0.486 & 0.423 \\
\bottomrule
\end{tabular}
\end{sc}
\end{small}
\end{table}

\textbf{Inference Settings.} Inference utilizes a 500-step iterative sampling procedure. To dynamically modulate generation diversity, a linear temperature annealing schedule is applied, where the temperature $T_t$ decays from $T_{\text{max}}=2.0$ to $T_{\text{min}}=1.0$ according to $T_t = T_{\text{min}} + (T_{\text{max}} - T_{\text{min}}) \cdot (1 - \frac{t}{N})$. Target sequence lengths are sampled uniformly from $U(200, 400)$.

\section{Implementation of Baselines}

For ProteoGAN, ProGen2, and CFP-Gen, we adopt the experimental results directly from the CFP-Gen publication \cite{cfpgen}, as our study utilizes the identical training and evaluation data splits. This ensures a fair and direct comparison with established benchmarks.

For Chroma and Pinal, which support natural language conditioning, we facilitate comparison by transforming the structured GO label combinations in the test set into natural language descriptions. Specifically, we construct input prompts by embedding the function name and target length into a standardized template. A representative prompt used for inference is:
\begin{quote}
    ``\textit{Generate a protein that functions as: 3-dehydroquinate dehydratase activity. The sequence length is approximately 285.}''
\end{quote}

For Ours(two-step) in the ablation study, we first train exclusively on structure tokens. During inference, the predicted structure tokens are passed to a pre-trained DPLM-2 inverse folding model to generate the corresponding sequence tokens.

\section{Semantic Explanations of Functional Constraints} \label{sec:go_semantic}

In this section, we provide detailed semantic definitions for the Gene Ontology (GO) term combinations utilized in our novelty assessment and case studies.

\subsection{Target Groups for Novelty and Diversity Testing}

To evaluate the model's capability to generate diverse structures within specific functional niches, we selected five distinct functional groups. As detailed in Table \ref{tab:target_groups}, these groups encompass a broad spectrum of biochemical activities, ranging from metal-sulfur cluster binding and electron transport to nucleic acid processing and enzymatic ligation.

\begin{table}[ht]
\centering
\caption{\textbf{Functional Target Groups.} Detailed breakdown of the GO term combinations used in the novelty and diversity experiments.}
\label{tab:target_groups}
\resizebox{\textwidth}{!}{
\begin{tabular}{@{}l p{0.18\textwidth} p{0.65\textwidth}@{}}
\toprule
\textbf{Group ID} & \textbf{GO Terms} & \textbf{Biological Semantics} \\
\midrule
\textbf{Iron-Sulfur Cluster} & GO:0004076, GO:0005506, GO:0051537, GO:0051539 & Involves the binding of iron ions and 2Fe-2S clusters, playing critical roles in electron transfer and catalytic processes. \\
\midrule
\textbf{Metallo-peptidase} & GO:0004477, GO:0004488 & Represents bifunctional enzymatic activities (methenyltetrahydrofolate cyclohydrolase and dehydrogenase) essential for the folate cycle and one-carbon metabolism. \\
\midrule
\textbf{NADH Dehydrogenase} & GO:0008137, GO:0048038, GO:0050136 & Encompasses NADH dehydrogenase (quinone/ubiquinone) activity, central to the mitochondrial electron transport chain and cellular respiration. \\
\midrule
\textbf{tRNA Ligase} & GO:0004070, GO:0016597 & Includes phosphopantothenoylcysteine decarboxylase and aminoacyl-tRNA ligase activities, fundamental for protein biosynthesis and coenzyme A metabolism. \\
\midrule
\textbf{RNA Binding} & GO:0003723, GO:0004523, GO:0030145 & Covers broad RNA binding capabilities and specific ribonuclease activities (e.g., RNA-DNA hybrid digestion), regulating gene expression and RNA stability. \\
\bottomrule
\end{tabular}
}
\end{table}

\subsection{Additional Case Study}



\begin{figure}[t]
    \centering
    \includegraphics[width=\linewidth]{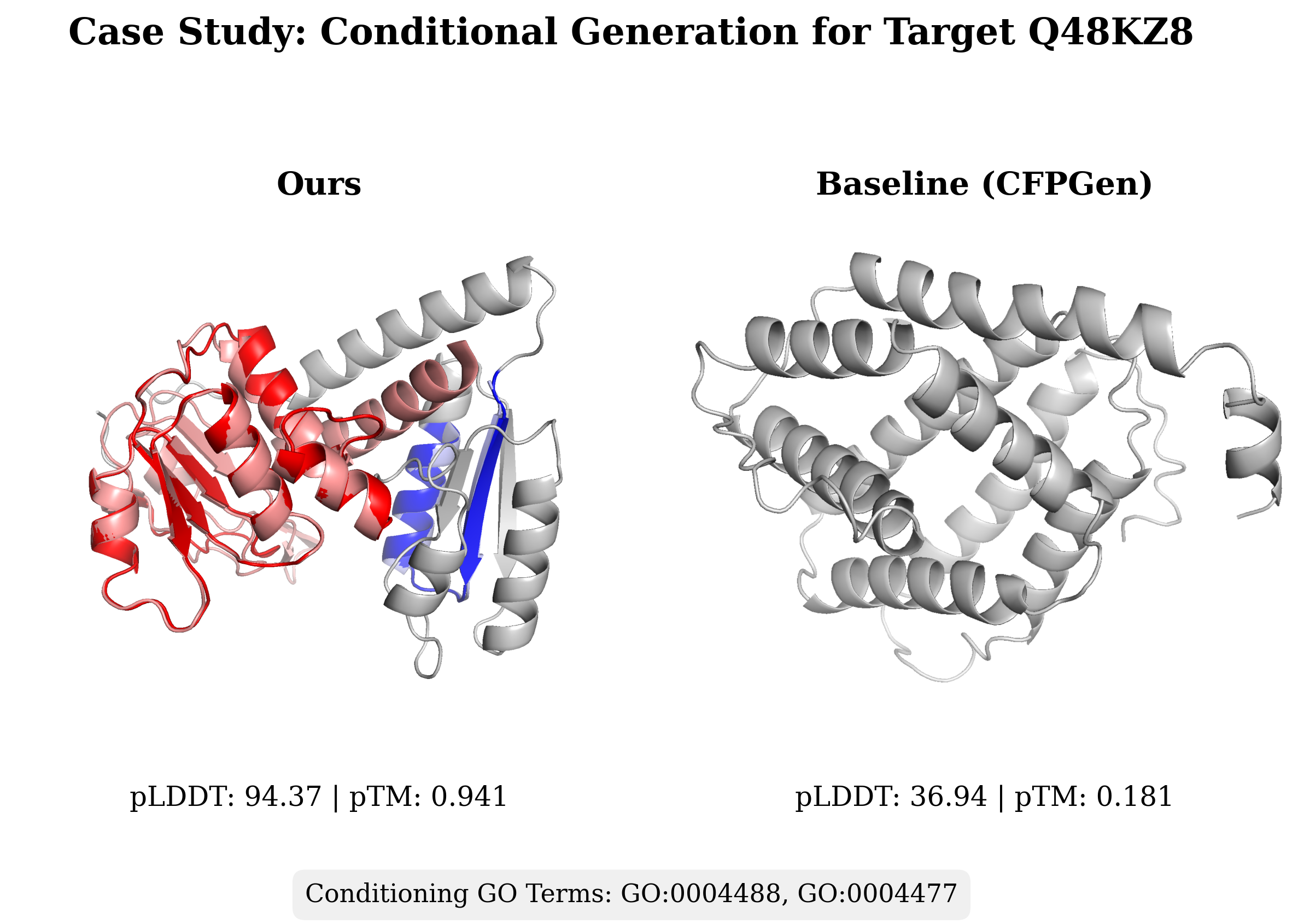}
    \caption{\textbf{Structural generation for Target Q48KZ8.} We compare the structures generated by our model (left) and the baseline (right) conditioned on dual functional constraints (GO:0004477, GO:0004488). The specific local motifs required for both functions is highlighted in red and blue, respectively.}
    \label{fig:case_study_train}
\end{figure}

We visualize another design case for Target Q48KZ8 (Fig.~\ref{fig:case_study_train}).
Structurally, our model successfully integrates the distinct functional motifs into a globally coherent and stable backbone (pLDDT: 94.37).
Sequentially, the generated protein exhibits 52.94\% sequence identity to the natural reference. It confirms that our model constructs distinct \textit{de novo} variants that preserve essential functional geometry.

\subsection{Functional Combinations for Case Studies}

We detail below the biological functions and constraints associated with the proteins examined in our case studies.

\paragraph{Case 1: Bifunctional Folate Enzyme.}
The first case study considers a bifunctional enzyme involved in folate metabolism, constrained by \texttt{GO:0004477} (methenyltetrahydrofolate cyclohydrolase activity) and \texttt{GO:0004488} (methylenetetrahydrofolate dehydrogenase (NADP+) activity). 
This functional combination requires the model to generate a protein structure capable of catalyzing two sequential biochemical reactions within the folate pathway, which typically entails a multi-domain organization or a structurally coordinated active site supporting both catalytic functions.

\paragraph{Case 2: Dehydrogenase with Cofactor Binding.}
The second case study focuses on a dehydrogenase that additionally requires explicit cofactor binding, specified by \texttt{GO:0008926} (mannitol-1-phosphate 5-dehydrogenase activity) and \texttt{GO:0051287} (NAD binding).
This functional specification imposes dual structural constraints: the generated protein must form a catalytically competent active site for the NAD(H)-dependent interconversion between D-fructose 6-phosphate and D-mannitol 1-phosphate, while simultaneously constructing a well-defined binding pocket to accommodate the NAD cofactor necessary for hydride transfer.


\end{document}